\documentclass[manuscript,screen]{acmart}
\setcopyright{none}
\renewcommand\footnotetextcopyrightpermission[1]{}

\AtBeginDocument{%
  }

\setcopyright{acmlicensed}
\copyrightyear{2026}
\acmYear{2026}
\acmDOI{XXXXXXX.XXXXXXX}

\acmJournal{TORS}
\acmVolume{37}
\acmNumber{4}
\acmArticle{111}
\acmMonth{8}

\usepackage{booktabs,multirow}
\usepackage{graphicx}
\usepackage[table]{xcolor}
\usepackage{enumitem}

\usepackage{amsthm}

\begin{document}

\title{Unpacking Provider Exposure Unfairness: A Diagnostic Framework Based on the Generalized Entropy Index}

\author{Jingzhi Li}
\affiliation{%
  \institution{Hefei University of Technology}
  \city{Hefei, Anhui}
  \country{China}}
\email{jingzhili826@gmail.com}

\author{Weijie Feng}
\affiliation{%
  \institution{Hefei University of Technology}
  \city{Hefei, Anhui}
  \country{China}}
\email{wjfeng@hfut.edu.cn}

\author{Ting Ran}
\affiliation{%
  \institution{Hefei University of Technology}
  \city{Hefei, Anhui}
  \country{China}}
\email{2025110498@mail.hfut.edu.cn}


\author{Zhiyong Cheng}
\affiliation{%
  \institution{Hefei University of Technology}
  \city{Hefei, Anhui}
  \country{China}
}
\email{jason.zy.cheng@gmail.com}

\author{Richang Hong}
\affiliation{%
  \institution{Hefei University of Technology}
  \city{Hefei, Anhui}
  \country{China}
}
\email{hongrc.hfut@gmail.com}

\renewcommand{\shortauthors}{Jingzhi Li et al.}

\begin{abstract}
Provider exposure fairness is essential to healthy multi-stakeholder recommender ecosystems. Yet existing evaluation practices often summarize exposure unfairness with a single score, making methods with different exposure distributions appear equally fair. Such scores cannot show whether method rankings remain stable as emphasis shifts from underexposed to highly exposed providers, or whether remaining unfairness lies mainly between or within provider groups. To address these limitations, we propose a diagnostic framework based on the \textbf{Generalized Entropy Index} (GEI), a family of inequality measures rooted in welfare economics. Its \textbf{tunable parameter}~$\alpha$ shifts emphasis across exposure levels, while its \textbf{exact decomposition} separates total unfairness into between-group and within-group components. The proposed framework leverages these properties through four diagnostic stages. \textbf{Configuration} specifies the values of~$\alpha$ and provider groupings. \textbf{Inequality profiling} traces unfairness across values of~$\alpha$, \textbf{ranking-reversal analysis} tests method-ranking stability, and \textbf{structural attribution} quantifies the between-group and within-group contributions. Together, these stages assess whether similar total unfairness scores imply stable method rankings and comparable structures. Experiments on three real-world datasets reveal multiple ranking reversals across values of~$\alpha$ and show that similar total unfairness can reflect either between-group disparity or within-group concentration.
\end{abstract}


\maketitle

\section{Introduction}
In recommender systems, provider exposure reflects how platforms allocate user attention across the providers that supply items. This allocation shapes the long-term structure of content ecosystems~\cite{Survey3,Survey4,chen1,Survey5}. When exposure consistently concentrates on a small subset of providers, users may encounter a narrower range of content, while smaller or emerging providers may receive fewer opportunities for growth~\cite{New-item1,New-item2,New-item3}. For example, on a music streaming platform, recommendations dominated by major-label artists may leave independent musicians with limited exposure, even when their content is comparable in quality~\cite{new1,new3,new6}. These consequences make provider exposure fairness a central evaluation concern in multi-stakeholder recommender systems.

A growing body of work has proposed in-processing~\cite{Ada2Fair,Multi-FR,FairDual} and post-processing~\cite{new8,new9,TFROM,CPFair,p-mmf,FairRec} methods to improve provider exposure fairness. Yet assessing the effectiveness of these interventions ultimately depends on how fairness is evaluated. Existing protocols often rely on a single fixed-sensitivity metric, such as the Gini coefficient~\cite{FairDual,XuCHMXR26} or Shannon entropy~\cite{fairdiverse,Post-hoc}, to summarize provider exposure unfairness. Such scores cannot reveal whether method rankings remain stable as the evaluation emphasis shifts from underexposed to highly exposed providers, or whether comparable levels of unfairness share the same structural composition. As a result, methods may appear equally fair even when their rankings reverse across sensitivity settings or their unfairness is driven by different structural sources. These limitations expose two diagnostic blind spots in existing provider exposure fairness evaluation, which we discuss next.

\begin{figure*}[!h]
\centering
\includegraphics[width=0.95\textwidth]{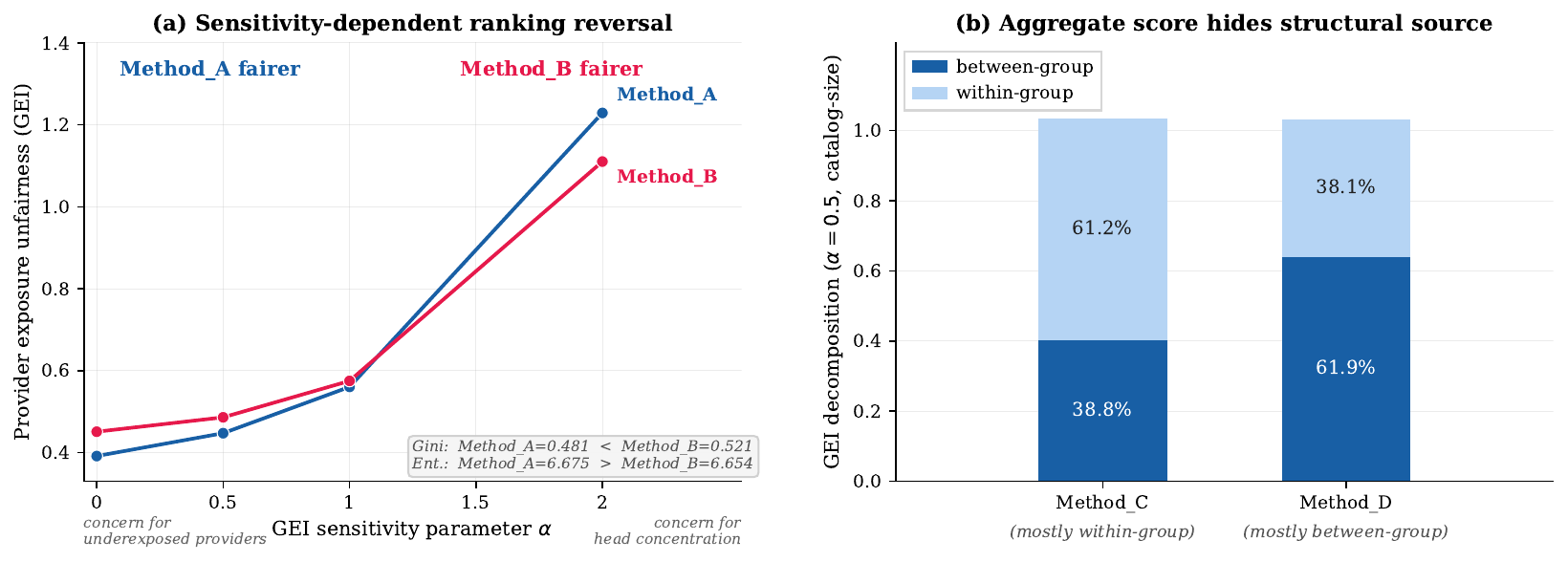}
\caption{%
Two diagnostic blind spots in provider exposure fairness evaluation.
\textbf{(a)}~Methods that appear similarly fair or follow one ranking at a given sensitivity setting may reverse their relative ordering as the Generalized Entropy Index (GEI) sensitivity parameter~$\alpha$ varies.
\textbf{(b)}~Methods with similar levels of total exposure unfairness may exhibit markedly different contributions from between-group and within-group unfairness, with unfairness dominated by either between-group disparity or within-group concentration.
}
\label{fig:motivation}
\end{figure*}

Figure~\ref{fig:motivation} illustrates the two diagnostic blind spots. The first blind spot is \textbf{fixed sensitivity}. Widely used metrics assign fixed weights to deviations across the exposure distribution, preventing evaluators from assessing whether fairness conclusions remain stable when different regions receive greater emphasis. As shown in Figure~\ref{fig:motivation}(a), Method~A may appear fairer than Method~B under a lower value of the GEI sensitivity parameter~$\alpha$. As~$\alpha$ increases and the evaluation places greater emphasis on highly exposed providers, their relative ordering may reverse, making Method~B the fairer choice. A single fixed-sensitivity score therefore cannot reveal whether method rankings remain stable as the evaluation emphasis shifts from underexposed to highly exposed providers.

The second blind spot is the \textbf{lack of structural attribution}. Providers naturally differ in structural attributes such as catalog size or interaction frequency. Exposure unfairness may therefore arise from disparities between provider groups, or from concentration within groups. As illustrated in Figure~\ref{fig:motivation}(b), Methods~C and~D may exhibit the same level of total exposure unfairness while having fundamentally different structural compositions. Method~C may be dominated by within-group concentration, whereas Method~D may be dominated by between-group disparity. A single aggregate exposure-unfairness score may make these methods appear equally fair despite their structural differences. Without structural attribution, evaluators cannot determine whether the remaining unfairness is driven primarily by between-group disparity or within-group concentration.

To overcome these limitations, we draw on welfare economics, where inequality measures are used not only to quantify disparities but also to identify their sources. Among these measures, GEI is particularly well suited to diagnostic evaluation because it combines a tunable sensitivity parameter~$\alpha$ with exact additive decomposition. Smaller values of~$\alpha$ emphasize exposure shortfalls among underexposed providers, whereas larger values emphasize exposure concentration among highly exposed providers. Meanwhile, additive decomposition separates total unfairness into between-group and within-group components without a residual term. Building on these properties, we develop a four-stage diagnostic protocol. The configuration stage specifies the sensitivity settings and provider groupings. The inequality profiling stage constructs multi-$\alpha$ inequality profiles to characterize exposure unfairness from different sensitivity perspectives. The ranking-reversal analysis stage identifies changes in the relative ordering of methods across sensitivity settings. Finally, the structural attribution stage quantifies the between-group and within-group contributions to total unfairness. Together, these stages reveal sensitivity-dependent fairness differences and the structural sources of exposure inequality, providing insights to guide the selection and design of fairness interventions.

We conduct extensive experiments on three real-world datasets. The results provide empirical evidence for both diagnostic limitations. Method rankings frequently reverse as the evaluation emphasis shifts from underexposed to highly exposed providers. Moreover, similar levels of total exposure unfairness can arise from markedly different structural compositions, dominated by either between-group disparity or within-group concentration. A case study further shows that the proposed diagnostic framework can lead to a different method choice from that suggested by conventional evaluation.

The main contributions of this work are summarized as follows:
\begin{itemize}[topsep=2pt,leftmargin=*]
\item We identify two diagnostic blind spots in existing provider exposure fairness evaluation: unstable method rankings across sensitivity settings and hidden structural sources of exposure unfairness.

\item We develop a four-stage diagnostic protocol based on the GEI that profiles unfairness across values of~$\alpha$, detects ranking reversals, and decomposes total exposure unfairness into between-group and within-group components.

\item We conduct extensive experiments on three real-world datasets, revealing frequent ranking reversals, distinct unfairness structures behind similar total scores, and method choices that differ from those suggested by conventional evaluation.
\end{itemize}

The remainder of this paper is organized as follows. Section~\ref{sec:related_work} reviews related work. Section~\ref{sec:method} presents the proposed diagnostic framework. Section~\ref{sec:experiments} describes the experimental setup and results. Section~\ref{Limitations} discusses limitations, and Section~\ref{conclu} concludes the paper.

\section{Related Work}
\label{sec:related_work}
\subsection{Provider Exposure Fairness in Multi-Stakeholder Recommendation}
\label{sec:rw_exposure}
Fairness in recommender systems concerns not only user utility but also how recommendation opportunities are distributed among items and providers. In ranking-based recommendation, item position strongly affects exposure, making a ranked list an allocation of limited user attention. Prior work has formalized this perspective by treating position-induced attention as a scarce resource. The Fairness of Exposure in Rankings~\cite{singh2018fairness} incorporates position-aware exposure into fairness constraints, while Equity of Attention~\cite{Equity_Attention} compares accumulated attention with accumulated relevance over time to support amortized fairness. These studies frame exposure fairness as an allocation problem rather than solely a prediction or ranking problem.

Provider exposure fairness is commonly studied within multi-stakeholder recommendation, where recommender systems simultaneously serve consumers, providers, and the platform~\cite{Multisided,Multi-stakeholder}. In this setting, exposure directly shapes providers' opportunities for content discovery and user engagement, making a fair allocation of exposure essential for maintaining a diverse and sustainable provider ecosystem. Yet recommender systems often disproportionately concentrate exposure among already advantaged providers, limiting the visibility and growth opportunities of long-tail or emerging providers~\cite{Unfairness,Survey3,landscape}.

Existing work defines provider exposure fairness through several normative principles. Proportional exposure~\cite{CPFair,Target1,new3} requires providers to receive exposure proportional to a merit signal, such as relevance or catalog size. Group-level parity~\cite{FairNeg,SDRO} seeks to reduce exposure disparities between predefined groups. Max-min fairness~\cite{p-mmf,LTP-MMF} prioritizes improvements for the worst-off providers. These notions differ in their units of comparison and reference standards. Consequently, fairness conclusions may depend not only on the recommendation method but also on the evaluative perspective adopted. This motivates evaluation approaches that examine whether method comparisons remain stable across perspectives and identify the sources of the observed unfairness.

\subsection{Fairness-Aware Methods for Provider Exposure}
\label{sec:rw_provider_methods}
A growing body of work has developed methods to improve provider exposure fairness by intervening at different stages of the recommendation pipeline. Depending on whether fairness considerations are introduced during model training or after recommendation scores have been generated, these approaches can be broadly divided into in-processing and post-processing methods.

In-processing methods incorporate fairness objectives directly into model training, thereby influencing the scores that determine how exposure is distributed in the final rankings. FairNeg~\cite{FairNeg} adjusts negative sampling to improve item-group fairness, while SDRO~\cite{SDRO} applies distributionally robust optimization to improve worst-case subgroup performance. Ada2Fair~\cite{Ada2Fair} addresses provider-side and consumer-side inequalities through adaptive sample weighting, and FairDual~\cite{FairDual} incorporates max-min fairness over provider-level exposure through dual optimization. Despite differences in their specific objectives, these methods all embed fairness considerations into the learning process. Their effects therefore depend on the recommendation architecture and fairness objective. Changes to these components may require the model to be retrained or the optimization procedure to be repeated.

Post-processing methods instead modify the ranked lists produced by a base recommender. A key practical advantage is that they are typically model-agnostic and can be deployed without changing the underlying recommendation model. FairRec~\cite{FairRec} formulates two-sided recommendation as an indivisible-goods allocation problem, in which recommendation opportunities are allocated while balancing consumer satisfaction and provider exposure. PMMF~\cite{p-mmf} treats positions in recommendation lists, together with the exposure they generate, as resources to be allocated and applies max-min fairness to prioritize disadvantaged providers. Subsequent work extends this allocation perspective to welfare optimization and long-term feedback dynamics~\cite{LTP-MMF}. Beyond allocation-based formulations, several post-processing studies interpret the trade-off between fairness and recommendation utility through economic concepts. TaxRank~\cite{TaxRank} treats accuracy loss on highly ranked items as a tax used to subsidize underexposed providers, while ElasticRank~\cite{ElasticRank} uses the concept of elasticity to characterize how much recommendation utility must be sacrificed to obtain a given improvement in fairness. Although post-processing methods are flexible, their effectiveness remains constrained by the quality, diversity, and score distribution of the candidate items supplied by the base recommender.

Because these methods pursue different fairness objectives and redistribute exposure through different mechanisms, a diagnostic evaluation that goes beyond a single aggregate score is needed to assess whether their fairness outcomes are stable and structurally comparable.

\subsection{Metrics for Exposure Fairness}
\label{sec:rw_metrics}
Evaluating provider exposure fairness first requires translating ranked recommendation outputs into provider-level exposure distributions. Existing work typically estimates item exposure using position-based discount models, which assign greater exposure to higher-ranked items~\cite{singh2018fairness,craswell2008experimental}, or browsing models that capture users' examination behavior~\cite{Equity_Attention,BernardB25}. Item-level exposure is then aggregated according to item ownership to obtain an exposure vector over providers. Once this vector is constructed, existing metrics can be broadly organized into three categories: group comparison metrics, distribution-level metrics, and target divergence metrics.

Group comparison metrics assess whether exposure differs across predefined provider groups. Measures such as exposure difference~\cite{new3,Equity_Attention} and min-max ratio~\cite{LinZZGLM17,KayaBT20} compare the average or total exposure received by different groups. These metrics are useful when fairness is defined in terms of disparities between protected and unprotected providers or between advantaged and disadvantaged groups. However, they do not by themselves characterize how exposure is distributed among providers within each group. A method may improve group-level parity while concentrating exposure on a small number of providers inside the groups. Group comparison metrics can therefore identify disparities between predefined groups, but they cannot determine how much of the overall unfairness originates between groups and how much remains within them.

Distribution-level metrics summarize the equality, diversity, or concentration of exposure across providers without requiring predefined groups. Common examples include the Gini coefficient~\cite{FuXGZHGXGSZM20,longterm}, Shannon entropy~\cite{fairdiverse,Survey5}, Jain's fairness index~\cite{SediqGSY13,LinZZGLM17}, and the Herfindahl-Hirschman Index~\cite{WangMLLZLLJM22,DhaliwalMS21}. These metrics compress the entire provider exposure distribution into a single scalar score, facilitating comparison across recommendation methods. However, each metric has a fixed mathematical form and a built-in sensitivity, meaning that it assigns different implicit importance to deviations in different regions of the exposure distribution. Comparisons based on a single metric may therefore reflect its particular sensitivity rather than a conclusion that remains stable when greater emphasis is placed on underexposed providers or on concentration among highly exposed providers. In conventional practice, testing ranking stability requires comparing distinct metrics that differ in scale, range, and underlying assumptions. This makes the resulting conclusions difficult to interpret consistently.

Unlike distribution-level metrics, which assess the shape of the observed exposure distribution, target divergence metrics evaluate fairness relative to an explicitly specified reference distribution. The target may be uniform or proportional to a provider merit signal, such as relevance or catalog size. Kullback-Leibler divergence~\cite{Survey3,Survey2}, Jensen-Shannon divergence~\cite{lin1991divergence}, and Generalized Cross Entropy~\cite{GCE} quantify the discrepancy between the observed and target exposure distributions. These metrics are flexible because the target can encode different normative assumptions about how exposure should be allocated. Their conclusions, however, may depend on how the target is specified, and different targets can yield different rankings for the same recommendation outputs. Moreover, a single divergence value cannot by itself reveal whether the observed deviation is primarily driven by between-group disparity or within-group concentration.

Overall, existing metrics are effective for quantifying particular aspects of provider exposure fairness. However, commonly used single-metric evaluation protocols do not provide a unified and internally consistent way to examine whether method rankings remain stable across sensitivity settings while also attributing total exposure unfairness to between-group and within-group components. Our framework addresses these two diagnostic needs within a single parameterized and decomposable formulation.

\subsection{Inequality Measures and GEI-based Fairness Evaluation}

Previous studies have developed flexible frameworks for fairness evaluation in recommender systems. Deldjoo et al.~\cite{deldjoo2021flexible} introduce a Generalized Cross Entropy-based framework that accommodates different fairness targets for users and items. Rampisela et al.~\cite{rampisela2024evaluation} systematically examine the theoretical and empirical properties of individual item exposure fairness measures, while Amigó et al.~\cite{amigo2023unifying} propose a unified information-theoretic framework covering diverse fairness definitions. These studies provide important foundations for fairness measurement, while also motivating further investigation into the diagnostic capabilities of inequality measures.

To support such diagnostic analysis, we draw on inequality measurement in welfare economics. Among existing inequality measures, the Generalized Entropy Index (GEI) provides a parameterized family of additively decomposable measures~\cite{Unified-Approach,SupeesunCMF24}. Its sensitivity parameter~$\alpha$ controls the relative emphasis placed on different regions of the exposure distribution, while its additive decomposition separates total inequality into between-group and within-group components without a residual term. These properties make GEI suitable for examining sensitivity-dependent fairness comparisons and the structural composition of provider exposure inequality.

GEI has also been applied to algorithmic fairness evaluation. Speicher et al.~\cite{Unified-Approach} employ generalized entropy measures to quantify unfairness in individual benefit distributions and examine individual- and group-level unfairness in supervised prediction. Building on these established foundations, we investigate their application to provider exposure fairness in recommender systems. Specifically, we develop a diagnostic protocol integrating multi-$\alpha$ inequality profiling, ranking-reversal analysis, and between-group and within-group decomposition. Our contribution lies in systematically combining these analytical properties for provider exposure diagnosis rather than introducing a new inequality measure.

\section{Method}
\label{sec:method}
We first define provider-level exposure and provider groupings, which serve as the basis for subsequent diagnostic analysis. We then introduce GEI as the diagnostic foundation and describe a four-stage protocol that supports inequality profiling, ranking-reversal analysis, and subgroup-based structural attribution.

\subsection{Problem Formulation}
\label{sec:problem}
We consider a recommender system with a set of users $\mathcal{U}$, a set of items $\mathcal{I}$, and a set of providers $\mathcal{P}$. Let $N_{\mathcal{U}}=|\mathcal{U}|$, $N_{\mathcal{I}}=|\mathcal{I}|$, and $M=|\mathcal{P}|$ denote the numbers of users, items, and providers, respectively. A provider is an entity that supplies items, such as a seller on an e-commerce platform or a creator on a video-sharing platform. We assume that each item is supplied by exactly one provider and represent this association using a mapping $\phi\colon\mathcal{I}\rightarrow\mathcal{P}$, where $\phi(i)=j$ indicates that item $i$ belongs to provider $j$. The catalog of provider $j$ is defined as $\mathcal{I}_j=\{i\in\mathcal{I}\mid\phi(i)=j\}$, and we assume that $|\mathcal{I}_j|>0$ for every $j\in\mathcal{P}$.

To support the subsequent structural decomposition of exposure unfairness, we partition providers according to a selected attribute, such as catalog size or interaction frequency. Let $g\colon\mathcal{P}\rightarrow\{1,\dots,C\}$ be a grouping function that assigns each provider to one of $C$ groups. For each $c\in\{1,\dots,C\}$, group $c$ and its size are defined as:
\begin{equation}
G_c=\{j\in\mathcal{P}\mid g(j)=c\},\quad M_c=|G_c|.
\end{equation}
We assume that $M_c>0$ for every group. The collection $\{G_1,\dots,G_C\}$ forms a partition of $\mathcal{P}$ and satisfies $\sum_{c=1}^{C}M_c=M$. This formulation assumes that group memberships are known and mutually exclusive. We discuss the limitations of this assumption in Section~\ref{Limitations}.

For each user $u\in\mathcal{U}$, let $R_u^K$ denote the generated top-$K$ recommendation list. For an item $i\in R_u^K$, its rank position is denoted by $r_{u,i}\in\{1,\dots,K\}$, where a smaller value indicates a higher position in the list. Because items at different ranking positions receive different amounts of user attention, we assign an exposure weight $w(r)>0$ to each position $r$. We assume that $w(r)$ is non-increasing in $r$, so an item ranked higher receives no less exposure than an item ranked lower. Following position-based exposure models~\cite{Equity_Attention,attention1,attention3}, we adopt the logarithmic discount function:
\begin{equation}
w(r)=\frac{1}{\log_2(r+1)}.
\label{eq:posi}
\end{equation}
The accumulated exposure received by item $i$ across all users is then defined as:
\begin{equation}
\label{eq:exposure}
x_i=\sum_{u\in\mathcal{U}:i\in R_u^K}w(r_{u,i}),
\end{equation}
where $x_i\geq 0$. When $w(r)=1$ for all positions, $x_i$ reduces to the number of users whose top-$K$ recommendation lists contain item $i$.
We use the average item exposure within a provider's catalog as the provider-level quantity of interest. Specifically, the exposure score of provider $j$ is defined as:
\begin{equation}
\label{eq:provider_exposure}
e_j=\frac{1}{|\mathcal{I}_j|}\sum_{i\in\mathcal{I}_j}x_i,\qquad j\in\mathcal{P}.
\end{equation}
Normalizing by $|\mathcal{I}_j|$ prevents catalog size alone from determining the provider-level score and allows providers to be compared in terms of the average exposure opportunity received by their items. Let $\mathbf{e}=(e_j)_{j\in\mathcal{P}}$ denote the provider exposure vector. Its mean is defined as:
\begin{equation}
\bar{e}=\frac{1}{M}\sum_{j\in\mathcal{P}}e_j.
\end{equation}
We assume that at least one recommended item receives positive exposure, so that $\bar{e}>0$.

GEI measures inequality in the provider-level values being analyzed. Its interpretation as unfairness depends on the normative exposure target. In our default setting, GEI is applied directly to $\mathbf{e}$, corresponding to a target of equal average item exposure across providers. Under this target, lower unfairness in $\mathbf{e}$ indicates a more equal allocation of average exposure opportunities. The framework is not restricted to this assumption. When each provider $j$ has a positive exposure target $t_j>0$, the same analysis can instead be applied to the exposure-to-target ratios $\{e_j/t_j\}_{j\in\mathcal{P}}$. Equality of these ratios indicates that providers receive exposure in the same proportion to their respective targets. 

\subsection{The Generalized Entropy Index}
\label{sec:gei}
Under the default target of equal average item exposure, diagnosing provider exposure unfairness requires measuring how unevenly exposure scores are distributed across providers. The GEI is suitable for this purpose because it quantifies distributional unfairness while allowing evaluators to vary which parts of the exposure distribution receive greater emphasis. We first define GEI for provider exposure and then introduce the two properties that support our diagnostic framework: tunable sensitivity and exact subgroup decomposition.

\paragraph{Definition.}
Given the provider exposure vector $\mathbf{e}$ and its mean $\bar{e}$ defined in Section~\ref{sec:problem}, GEI~\cite{cowell1980structure,shorrocks1980class,Unified-Approach} measures provider exposure unfairness as:
\begin{equation}
\label{eq:gei}
\mathrm{GEI}(\alpha) =
\begin{cases}
\displaystyle \frac{1}{M\,\alpha(\alpha - 1)}
  \sum_{j=1}^{M} \left[
  \left( \frac{e_j}{\bar{e}} \right)^{\!\alpha}
  - 1 \right], & \alpha \neq 0,1, \\[12pt]
\displaystyle \frac{1}{M} \sum_{j=1}^{M}
  \frac{e_j}{\bar{e}} \ln \frac{e_j}{\bar{e}},
  & \alpha = 1, \\[12pt]
\displaystyle \frac{1}{M} \sum_{j=1}^{M}
  \ln \frac{\bar{e}}{e_j}, & \alpha = 0.
\end{cases}
\end{equation}
The ratio $e_j/\bar{e}$ represents the exposure score of provider $j$ relative to the system-wide mean. A ratio above one indicates above-average exposure, whereas a ratio below one indicates below-average exposure. GEI summarizes the dispersion of these relative exposure ratios across providers. Under the equal-average-exposure target, a larger GEI value indicates greater exposure unfairness, and $\mathrm{GEI}(\alpha)=0$ if and only if all providers receive the same exposure score. The cases $\alpha=1$ and $\alpha=0$ are obtained as continuous limits of the general form and correspond to the Theil~$T$ index and the Mean Log Deviation (MLD), respectively. Because GEI is computed from exposure relative to the mean, it measures the shape of the exposure distribution rather than its absolute scale. At a fixed $\alpha$, GEI values can be compared across recommendation methods evaluated under the same exposure definition and dataset. However, because each value of $\alpha$ induces a different functional form and scale, raw GEI magnitudes at different values of $\alpha$ are not directly comparable.

\paragraph{Zero-exposure handling.}
Some providers may receive zero exposure because none of their items appears in the generated top-$K$ recommendation lists. Zero values create undefined terms for some GEI settings. In particular, $\alpha=0$ contains the logarithmic term $\ln(\bar{e}/e_j)$, while negative values of $\alpha$ involve negative powers of $e_j/\bar{e}$. Both cases require strictly positive exposure values.

To retain zero-exposure providers while avoiding undefined numerical operations, we apply a small scale-relative smoothing term before computing GEI:
\begin{equation}
\label{eq:smoothing}
\epsilon=\eta\bar{e},
\qquad
\tilde{e}_j=e_j+\epsilon,
\qquad
\tilde{\bar{e}}=
\frac{1}{M}
\sum_{j=1}^{M}\tilde{e}_j,
\end{equation}
where $\eta>0$ is a small constant. GEI is then computed by replacing $e_j/\bar{e}$ in Eq.~\eqref{eq:gei} with $\tilde{e}_j/\tilde{\bar{e}}$. Because $\epsilon$ is proportional to the original mean exposure, the smoothing procedure preserves scale invariance: multiplying all exposure scores by the same positive constant does not change the resulting normalized ratios. In our implementation, we set $\eta=10^{-12}$.

Smoothing is used only for the numerical computation of GEI and does not alter the original definition of provider exposure. For any fixed $\alpha>0$, the contribution of a zero-exposure provider remains finite as $\eta\rightarrow0$. At $\alpha=0$, the contribution of a smoothed zero-exposure provider increases logarithmically as $\eta$ decreases, making the MLD sensitive to both the smoothing value and the proportion of zero-exposure providers. For $\alpha<0$, the corresponding contribution diverges as a power of $1/\eta$, so the resulting GEI may be dominated by the smoothing choice when zero-exposure providers are common. We therefore exclude negative values of $\alpha$ from the default sensitivity range and retain $\alpha=0$ as the boundary case with the strongest emphasis on underexposed providers.

For notational simplicity, the following definitions are written in terms of $e_j$. When smoothing is required, $e_j$ and all derived group and system means are replaced by their smoothed counterparts.

\paragraph{Tunable sensitivity.}
Let $s_j=e_j/\bar{e}$ denote the relative exposure ratio of provider $j$. The parameter $\alpha$ controls how GEI responds to differences in these ratios. Smaller values of $\alpha$ place greater emphasis on exposure shortfalls among underexposed providers ($s_j<1$), whereas larger values are more sensitive to exposure concentration among highly exposed providers ($s_j>1$). Specifically, $\alpha=0$ emphasizes severe underexposure, $\alpha=0.5$ provides a moderate lower-tail perspective, $\alpha=1$ serves as a conventional reference, and $\alpha=2$ emphasizes high-exposure concentration. However, $\alpha=1$ is not universally neutral, and no single value of $\alpha$ provides a balanced evaluation for all fairness concerns. These values represent different sensitivity perspectives rather than fixed fairness thresholds. Therefore, the choice of $\alpha$ should reflect the fairness concern being evaluated. Examining method rankings across different values of $\alpha$ further reveals whether fairness conclusions remain stable as the evaluation emphasis changes.

\paragraph{Subgroup decomposition.}
A single GEI value quantifies total provider exposure unfairness but does not describe its structural composition. Exposure unfairness may reflect systematic differences in average exposure across provider groups, unequal exposure among providers within the same group, or a combination of both. The between-group component asks whether predefined groups receive different mean exposure scores, whereas the within-group component asks whether providers belonging to the same group receive unequal exposure. The provider partition introduced in Section~\ref{sec:problem} allows these components to be distinguished. GEI admits an exact additive decomposition with no residual term~\cite{bourguignon1979decomposable,shorrocks1980class}:
\begin{equation}
\label{eq:decomp}
  \mathrm{GEI}_{\mathrm{total}}(\alpha)
  \;=\;
  \mathrm{GEI}_{\mathrm{between}}(\alpha)
  \;+\;
  \mathrm{GEI}_{\mathrm{within}}(\alpha).
\end{equation}
Because the decomposition is residual-free, the between-group and within-group components fully account for total exposure unfairness under the specified provider partition.

For each group $G_c$, let its mean provider exposure score be $\bar{e}_c=
\frac{1}{M_c}
\sum_{j\in G_c}e_j$.
The between-group component measures unfairness arising from
differences in these group means:
\begin{equation}
\label{eq:between}
  \mathrm{GEI}_{\mathrm{between}}(\alpha)
  \;=\;
  \begin{cases}
  \displaystyle
  \frac{1}{M\,\alpha(\alpha-1)}
  \sum_{c=1}^{C} M_c
  \left[
    \left(\frac{\bar{e}_c}{\bar{e}}\right)^{\!\alpha} - 1
  \right],
  & \alpha \neq 0,1, \\[12pt]
  \displaystyle
  \sum_{c=1}^{C} \frac{M_c}{M}\,
  \frac{\bar{e}_c}{\bar{e}}\,
  \ln \frac{\bar{e}_c}{\bar{e}},
  & \alpha = 1, \\[12pt]
  \displaystyle
  \sum_{c=1}^{C} \frac{M_c}{M}\,
  \ln \frac{\bar{e}}{\bar{e}_c},
  & \alpha = 0.
  \end{cases}
\end{equation}
The ratio $\bar{e}_c/\bar{e}$ compares the mean exposure of group $G_c$ with the system-wide mean. The between-group component can therefore be interpreted as the unfairness that would remain if every provider within group $G_c$ received the same exposure score $\bar{e}_c$. A large between-group value indicates that differences in average exposure across groups account for a substantial portion of total unfairness. The within-group component is defined as:
\begin{equation}
\label{eq:within}
  \mathrm{GEI}_{\mathrm{within}}(\alpha)
  \;=\;
  \sum_{c=1}^{C}
  \underbrace{
    \frac{M_c}{M}
    \left(\frac{\bar{e}_c}{\bar{e}}\right)^{\!\alpha}
  }_{\omega_c}
  \;\mathrm{GEI}_c(\alpha),
\end{equation}
where $\mathrm{GEI}_c(\alpha)$ is computed using only the exposure scores of providers in group $G_c$. Thus, $\mathrm{GEI}_c(\alpha)$ measures how unevenly exposure is distributed among providers within that group.

The weight $\omega_c$ determines the contribution of group $G_c$ to the overall within-group component. It depends on the group's share of providers and its mean exposure relative to the system-wide mean. When $\alpha=0$, the weight reduces to $M_c/M$, so each group is weighted only by its share of providers. When $\alpha=1$, the weight becomes $\frac{M_c}{M}\frac{\bar{e}_c}{\bar{e}}$, so the contribution also depends on the group's relative mean exposure. Larger values of $\alpha$ increase the relative contribution of groups with higher mean exposure. As $\alpha$ decreases toward zero, this additional mean-exposure weighting weakens and disappears at $\alpha=0$. The lower-tail sensitivity of $\alpha=0$ instead operates through the group-specific term $\mathrm{GEI}_c(0)$, which emphasizes exposure shortfalls among providers within each group. A large within-group component therefore indicates substantial exposure unfairness among providers that share the same grouping attribute. The decomposition describes a statistical partition of total unfairness under the specified provider grouping; it does not identify the causal mechanisms that generated the observed exposure distribution.

\paragraph{Relationships with Conventional Fairness Metrics}
GEI has direct mathematical relationships with several conventional fairness metrics. For a fixed number of providers, GEI(1) is a monotonically decreasing transformation of Shannon entropy. Similarly, $\mathrm{GEI}(2)$ is a monotonic transformation of the Herfindahl--Hirschman Index (HHI) and Jain's fairness index. These relationships imply that the corresponding metrics produce equivalent fairness rankings when evaluated on the same provider exposure distributions with a fixed number of providers, after accounting for their evaluation directions. However, these mathematical equivalences do not imply that a single metric is sufficient for fairness diagnosis. Conventional metrics such as Gini, Shannon entropy, HHI, and Jain's fairness index each capture particular aspects of exposure inequality. When used individually, they do not directly reveal whether method rankings change across different sensitivity settings or how total unfairness is distributed between and within provider groups.

\subsection{Diagnostic Analysis Protocol}
\label{sec:diagnostic}
\begin{figure*}[!h]
\centering
\includegraphics[width=\textwidth]{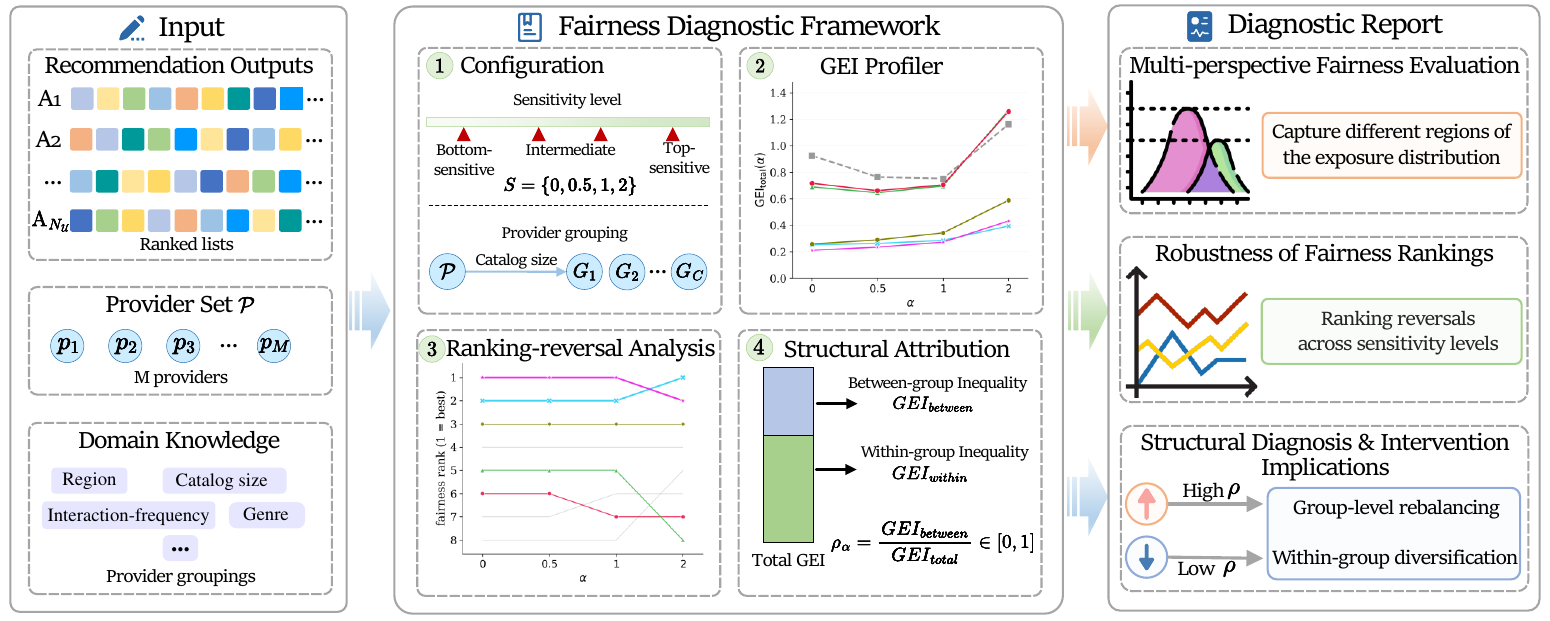}
\caption{Overview of the proposed four-stage diagnostic protocol for provider exposure fairness evaluation.}
\label{fig:overview}
\end{figure*}

The proposed protocol moves beyond a single unfairness score to examine whether comparative fairness conclusions remain stable across sensitivity settings and how total exposure unfairness is divided into between-group and within-group components. Figure~\ref{fig:overview} summarizes the four-stage diagnostic protocol.

The protocol takes as input the recommendation outputs of $N$ candidate methods $\{\mathcal{A}_1,\dots,\mathcal{A}_N\}$, the provider set $\mathcal{P}$, and one or more provider partitions. For each method, the recommendation outputs are converted into item-level exposure scores and then aggregated into provider exposure scores according to the definitions in Section~\ref{sec:problem}. The protocol produces sensitivity-indexed inequality profiles, ranking-reversal results, and partition-specific structural attribution results.

The protocol begins with configuration (Stage 1), which specifies the sensitivity set $\mathcal{S}$ and provider groupings. In this study, we use $\mathcal{S}=\{0,0.5,1,2\}$ to cover different sensitivity perspectives, as discussed in Section~3.2. Provider groupings define the partitions used for structural attribution. Both the sensitivity set and provider groupings can be adjusted according to the evaluation objectives.

Stage~2, \textbf{inequality profiling}, records the exposure unfairness of each method across the selected sensitivity settings. Method $\mathcal{A}_n$ is represented by the sensitivity-indexed profile $\{\mathrm{GEI}_n(\alpha)\}_{\alpha\in\mathcal{S}}$. Because raw GEI magnitudes are not directly comparable across different values of $\alpha$, each profile point is interpreted relative to other methods evaluated at the same sensitivity level. The profiles provide a unified view of method performance across $\mathcal{S}$ and make changes in relative standing visible. However, changes in a single method's raw GEI values across $\alpha$ do not reflect changes in the amount of unfairness, because different values of $\alpha$ induce different scales.

Stage~3, \textbf{ranking-reversal analysis}, examines whether the fairness ordering of methods remains stable across values of $\alpha$. Because a lower GEI value indicates lower exposure unfairness under the stated target, methods are ranked in ascending order of $\mathrm{GEI}_n(\alpha)$ at each sensitivity level. For two methods $\mathcal{A}_n$ and $\mathcal{A}_m$, a ranking reversal occurs if there exist $\alpha_1,\alpha_2\in\mathcal{S}$ such that
\begin{equation}
    \mathrm{GEI}_n(\alpha_1)<\mathrm{GEI}_m(\alpha_1)
\quad \text{and} \quad
\mathrm{GEI}_n(\alpha_2)>\mathrm{GEI}_m(\alpha_2).
\end{equation}
In the empirical analysis, differences within a predefined numerical tolerance are treated as ties and are not counted as ranking reversals. The presence of ranking reversals indicates that the comparative conclusion between two methods depends on the selected sensitivity setting.

Stage~4 (\textbf{structural attribution}) characterizes how total exposure unfairness is divided into between-group and within-group components. For each method and each $\alpha\in\mathcal{S}$, we decompose total unfairness using Eqs.~\eqref{eq:decomp}--\eqref{eq:within} and compute the \emph{diagnostic ratio}:
\begin{equation}
\label{eq:rho}
\rho_n(\alpha)
=
\frac{\mathrm{GEI}_{n,\mathrm{between}}(\alpha)}
{\mathrm{GEI}_{n,\mathrm{total}}(\alpha)}.
\end{equation}
We compute $\rho_n(\alpha)$ only when $\mathrm{GEI}_{n,\mathrm{total}}(\alpha)>0$. When total unfairness is zero, both the between-group and within-group components are reported as zero, and the ratio is omitted because there is no unfairness to attribute. For positive total unfairness, $\rho_n(\alpha)\in[0,1]$. A value of $\rho_n(\alpha)=1$ indicates that all measured unfairness is attributed to differences in mean exposure across provider groups, whereas $\rho_n(\alpha)=0$ indicates that all measured unfairness is attributed to unequal exposure among providers within the groups. We interpret $\rho_n(\alpha)$ comparatively rather than using a fixed threshold. A value of 0.5 serves as a descriptive reference indicating equal contributions from the two components, rather than a threshold for determining fairness. A larger value indicates that between-group disparity accounts for a greater share of total unfairness, whereas a smaller value indicates a greater contribution from within-group concentration. The diagnostic ratio may also vary across $\mathcal{S}$. For example, if $\rho_n(\alpha)$ is larger at smaller values of $\alpha$ but lower at larger values, the relative composition of unfairness changes as the evaluation shifts from lower-tail to upper-tail sensitivity. Such a pattern indicates that both components should be reported across sensitivity settings rather than summarized by a single structural characterization. When multiple provider groupings are considered, similar $\rho_n(\alpha)$ patterns across groupings may indicate that the structural interpretation is robust to the choice of partition. Divergent patterns simply indicate that the attribution depends on the structural attribute used to group providers.

Overall, the protocol transforms provider exposure fairness evaluation from a single-score comparison into a multi-perspective diagnostic process. Inequality profiling organizes method performance across sensitivity settings, ranking-reversal analysis tests the stability of comparative conclusions, and structural attribution quantifies the relative contributions of between-group disparity and within-group concentration under each provider partition.

\section{Experiments}
\label{sec:experiments}
We organize the experiments around four research questions that progress from metric comparison to diagnostic analysis and practical implications. We additionally conduct robustness experiments to assess the stability and reliability of our findings.
\begin{itemize}[topsep=2pt,leftmargin=*]
\item \textbf{RQ1.} How do GEI-based evaluations relate to conventional exposure-fairness metrics, and what complementary information does GEI provide?
 
\item \textbf{RQ2.} Does multi-$\alpha$ analysis reveal sensitivity-dependent fairness patterns and ranking instabilities that single fixed-sensitivity scores fail to capture?
 
\item \textbf{RQ3.} How is provider exposure unfairness divided into between-group and within-group components across different methods and provider partitions?
 
\item \textbf{RQ4.} How can the diagnostic findings inform the selection and design of provider exposure fairness interventions?
\end{itemize}

\subsection{Experimental Settings}
\subsubsection{Datasets}
\label{sec:datasets}
We conduct experiments on three publicly available recommendation datasets: Amazon Video \& Games, Steam, and BeerAdvocate. These datasets cover e-commerce, digital gaming, and beer review domains, respectively. Their item metadata allows us to construct a single-valued item-to-provider mapping for provider exposure fairness evaluation. Table~\ref{tab:dataset} summarizes the statistics of the processed datasets.
\begin{table}[!h]
\centering
\small
\caption{Dataset statistics after preprocessing.}
\label{tab:dataset}
\begin{tabular}{lrrrrr}
\toprule
\textbf{Dataset} & \textbf{\#Users} & \textbf{\#Items} &
\textbf{\#Inter.} & \textbf{\#Prov.} & \textbf{Sparsity} \\
\midrule
Steam & 8,183 & 1,196 &
86,385 & 42 &
99.12\% \\
Video \& Games & 42,003 &
13,912 & 382,954 &
159 & 99.93\% \\
BeerAdvocate &
14,172 & 18,678 &
1,373,246 & 637 &
99.48\% \\
\bottomrule
\end{tabular}
\end{table}

Amazon Video \& Games\footnote{\url{https://amazon-reviews-2023.github.io/}} contains user reviews from Amazon's Video \& Games category. Each product is treated as an item, and its brand is treated as the provider. Ratings greater than 3 are regarded as positive interactions. BeerAdvocate\footnote{\url{https://cseweb.ucsd.edu/~jmcauley/datasets.html}} contains user reviews of beers from different breweries. Each beer is treated as an item, and its associated brewery is treated as the provider. This dataset provides explicit item-to-provider associations, allowing us to evaluate provider exposure fairness using actual provider identities. Reviews with an overall rating \(\ge 3\) are treated as positive interactions. Steam\footnotemark[\value{footnote}] contains user play records from the Steam gaming platform. Each game is treated as an item, and its publisher is treated as the provider. Play records exceeding 10 hours are regarded as positive interactions.

For all datasets, we apply 5-core filtering to users and items to remove extremely sparse interaction records. Following the preprocessing strategy supported by FairDiverse [48], we merge providers with fewer than 10 retained items into a single auxiliary provider. This aggregation reduces the number of sparsely represented provider units, and the same provider mapping is used for all evaluated methods to ensure consistent fairness comparisons. Specifically, 114, 52, and 614 providers are merged on Amazon Video \& Games, Steam, and BeerAdvocate, respectively, accounting for 5.24\%, 27.26\%, and 21.52\% of retained items and 3.00\%, 21.41\%, and 9.90\% of retained interactions, respectively. For each user, interactions are chronologically divided into training, validation, and test sets at a ratio of 8:1:1.

\subsubsection{Provider Groupings}
\label{sec:groupings}
The subgroup decomposition in RQ3 requires a provider partition. We construct provider groups along two complementary dimensions to examine how structural attribution depends on the grouping criterion. The first dimension is \textbf{catalog size}, defined as the number of retained items supplied by a provider. According to their catalog sizes, providers are divided into small-, medium-, and large-catalog groups based on the terciles of the catalog-size distribution. This partition captures supply-side heterogeneity and examines whether groups with different catalog scales receive systematically different mean exposure scores. The second dimension is \textbf{interaction frequency}, defined as the total number of training interactions associated with a provider's items divided by the number of items supplied by that provider. We use the corresponding 33rd and 67th percentiles to divide providers into low-, medium-, and high-interaction groups. This partition captures differences in average item-level demand and examines whether providers with different demand intensities exhibit different structural compositions of exposure unfairness. Providers whose values equal a percentile boundary are assigned to the higher group. For Amazon Video \& Games, Steam, and BeerAdvocate, the auxiliary provider is assigned to the large-catalog group and the medium-interaction group under the two respective grouping schemes.

\subsubsection{Evaluated Methods}
\label{sec:methods}
To assess the generality of the proposed diagnostic framework across recommendation architectures and fairness intervention strategies, we consider two backbone recommenders and eight representative provider-fairness methods. The backbone recommenders are LightGCN~\cite{LightGCN}, a graph-based collaborative filtering model, and SASRec~\cite{sasrec}, a sequential recommendation model. Both are evaluated in their standard forms without fairness interventions and serve as accuracy-oriented baselines.
 
The fairness-aware methods cover two intervention paradigms. In-processing methods incorporate fairness objectives or corrections during model training. This category includes IPS~\cite{IPSS}, SDRO~\cite{SDRO}, FOCF~\cite{focf}, and FairDual~\cite{FairDual}. Post-processing methods improve provider fairness by re-ranking the recommendation lists produced by a backbone recommender. This category includes FairRec~\cite{FairRec}, CPFair~\cite{CPFair}, PMMF~\cite{p-mmf}, and ElasticRank~\cite{ElasticRank}. These methods allow us to examine how the diagnostic findings vary across recommendation architectures and intervention paradigms.
 
\subsubsection{Evaluation Metrics}
\label{sec:metrics}
We evaluate each method in terms of recommendation accuracy and provider exposure fairness. For recommendation accuracy, we report NDCG@$K$, HR@$K$, and MRR@$K$. Higher values indicate better recommendation performance.
 
All provider fairness metrics are computed from the same provider-level exposure score vector defined in Section~\ref{sec:problem}. We include the Gini coefficient (Gini) and Shannon entropy (Ent.) as conventional single-score references~\cite{FairDual,fairdiverse}. For the diagnostic evaluation, we report the GEI family at $\mathcal{S}=\{0,0.5,1,2\}$, denoted as GEI(0), GEI(0.5), GEI(1), and GEI(2), respectively. Lower GEI values indicate lower exposure unfairness under the specified sensitivity setting.
 
\subsubsection{Implementation Details}
\label{sec:implementation}
All experiments are implemented in PyTorch and conducted on a single NVIDIA RTX 3090 GPU. For the backbone recommenders, the embedding dimension is set to 64, and Adam is used for optimization. The learning rate is selected from $\{10^{-1},10^{-2},10^{-3},10^{-4},10^{-5}\}$. For GNN-based models, the number of graph convolutional layers is additionally selected from $\{1,2,3\}$. The fairness-aware baselines are implemented based on the original code released by their authors and the FairDiverse toolkit~\cite{fairdiverse}. The search ranges of method-specific hyperparameters follow the corresponding original papers, and all configurations are evaluated on the validation set. We adopt full-ranking evaluation for all methods. For each user, recommendation scores are computed over the entire item catalog, excluding items with which the user interacted during training, to generate the top-$K$ recommendation list. Recommendation accuracy is evaluated against the held-out positive interactions, while provider exposure fairness is computed from the resulting top-$K$ lists. Fairness-aware methods may achieve lower unfairness at the cost of recommendation accuracy. To mitigate this confounding effect, we select the configuration whose validation NDCG@$K$ is closest to $90\%$ of the corresponding backbone's validation performance. This controlled-accuracy protocol aims to reduce accuracy differences across methods. All main experiments are repeated with five different random seeds, and the reported results are averaged across these runs. Unless otherwise specified, the recommendation-list length is set to $K=20$. The experimental code and datasets are available at \url{https://github.com/Tam-JQK/GEI-Based}.

\begin{table*}[!h]
\centering
\caption{Overall results with LightGCN as the backbone. The best value in each row is in \textbf{bold} and the second-best is \underline{underlined}. The backbone serves as an accuracy reference. $\downarrow$ indicates lower is better and $\uparrow$ higher is better.}
\label{tab:overall_lightgcn}
\small
\setlength\tabcolsep{1.0pt}
\renewcommand{\arraystretch}{0.7}
\begin{tabular}{cc c cccc cccc}
\toprule
& & \multicolumn{1}{c}{Backbone} & \multicolumn{4}{c}{In-Processing} & \multicolumn{4}{c}{Post-Processing} \\
\cmidrule(lr){3-3} \cmidrule(lr){4-7} \cmidrule(lr){8-11}
Dataset & Metric & LightGCN & SDRO & IPS & FOCF & FairDual & FairRec & CPFair & PMMF & ElasticRank \\
\midrule
\multirow{9}{*}{Steam} & NDCG@20 & \textbf{0.1591} & 0.1428 & 0.1416 & 0.1421 & \underline{0.1457} & 0.1409 & 0.1407 & 0.1441 & 0.1442 \\
 & HR@20 & \textbf{0.3614} & 0.3237 & 0.2989 & 0.3326 & 0.3317 & \underline{0.3586} & 0.2745 & 0.3072 & 0.2860 \\
 & MRR@20 & 0.1126 & 0.1015 & 0.1072 & 0.0987 & 0.1036 & 0.0881 & \underline{0.1137} & 0.1076 & \textbf{0.1149} \\
\cmidrule(lr){2-11}
 & Gini$\downarrow$ & 0.8677 & 0.6676 & 0.6728 & 0.7417 & 0.8231 & 0.7950 & \textbf{0.5443} & 0.6168 & \underline{0.6093} \\
 & Ent.$\uparrow$ & 2.8749 & 4.1606 & 4.1791 & 3.8164 & 3.2757 & 3.3906 & \textbf{4.6126} & 4.2231 & \underline{4.4053} \\
\cmidrule(lr){2-11}
 & GEI(0)$\downarrow$ & 4.8393 & 1.0728 & 0.9793 & 1.4220 & 2.1233 & 1.5895 & \textbf{0.5561} & \underline{0.7063} & 0.8876 \\
 & GEI(0.5)$\downarrow$ & 2.0100 & 0.8646 & 0.8417 & 1.1176 & 1.5046 & 1.3425 & \textbf{0.5176} & 0.7064 & \underline{0.7021} \\
 & GEI(1)$\downarrow$ & 1.7449 & 0.8538 & 0.8409 & 1.0923 & 1.4671 & 1.3875 & \textbf{0.5405} & 0.8104 & \underline{0.6841} \\
 & GEI(2)$\downarrow$ & 3.1883 & 1.3253 & 1.2329 & 1.7405 & 2.7497 & 2.5530 & \textbf{0.7971} & 1.5610 & \underline{0.9813} \\
\midrule
\multirow{9}{*}{Video \& Games} & NDCG@20 & \textbf{0.0219} & 0.0198 & 0.0197 & 0.0196 & 0.0197 & 0.0197 & \underline{0.0199} & 0.0197 & 0.0194 \\
 & HR@20 & \textbf{0.0540} & 0.0521 & 0.0481 & 0.0504 & 0.0490 & \underline{0.0533} & 0.0475 & 0.0464 & 0.0450 \\
 & MRR@20 & \textbf{0.0145} & 0.0122 & 0.0128 & 0.0122 & 0.0128 & 0.0116 & 0.0135 & \underline{0.0136} & 0.0135 \\
\cmidrule(lr){2-11}
 & Gini$\downarrow$ & 0.9178 & 0.7109 & 0.7619 & 0.8943 & 0.9018 & 0.7414 & \underline{0.4492} & \textbf{0.3777} & 0.5285 \\
 & Ent.$\uparrow$ & 3.8069 & 5.9181 & 5.3995 & 4.3682 & 4.1230 & 4.9674 & \underline{6.6381} & \textbf{6.9398} & 6.6126 \\
\cmidrule(lr){2-11}
 & GEI(0)$\downarrow$ & 9.4704 & 3.7917 & 2.1088 & 13.2171 & 7.7302 & 1.0822 & \underline{0.3468} & \textbf{0.3047} & 0.5786 \\
 & GEI(0.5)$\downarrow$ & 2.2185 & 1.1273 & 1.2135 & 2.1303 & 2.0496 & 1.1766 & \underline{0.3790} & \textbf{0.2660} & 0.4965 \\
 & GEI(1)$\downarrow$ & 2.4302 & 0.9668 & 1.3263 & 2.0411 & 2.2110 & 1.6257 & \underline{0.4677} & \textbf{0.2586} & 0.4854 \\
 & GEI(2)$\downarrow$ & 13.0853 & 1.3810 & 4.0138 & 6.6458 & 9.5891 & 8.3282 & 1.1191 & \textbf{0.3453} & \underline{0.6657} \\
\midrule
\multirow{9}{*}{BeerAdvocate} & NDCG@20 & \textbf{0.0359} & 0.0319 & 0.0325 & 0.0326 & 0.0325 & 0.0309 & \underline{0.0327} & 0.0323 & 0.0323 \\
 & HR@20 & \textbf{0.2106} & 0.1997 & 0.1967 & 0.1986 & 0.1995 & \underline{0.2049} & 0.2000 & 0.2014 & 0.2000 \\
 & MRR@20 & \textbf{0.0563} & 0.0506 & 0.0492 & 0.0530 & 0.0524 & 0.0398 & \underline{0.0556} & 0.0544 & 0.0553 \\
\cmidrule(lr){2-11}
 & Gini$\downarrow$ & 0.9720 & \underline{0.8601} & 0.9619 & 0.9529 & 0.9674 & \textbf{0.8066} & 0.9161 & 0.9077 & 0.9021 \\
 & Ent.$\uparrow$ & 4.5215 & \textbf{6.6110} & 4.8198 & 5.1954 & 4.2384 & 6.0239 & 6.0464 & 6.1663 & \underline{6.2050} \\
\cmidrule(lr){2-11}
 & GEI(0)$\downarrow$ & 19.3036 & \underline{3.9367} & 13.4916 & 18.2011 & 21.0112 & \textbf{1.4421} & 12.2051 & 11.7236 & 10.8118 \\
 & GEI(0.5)$\downarrow$ & 2.9349 & \underline{1.6647} & 2.6880 & 2.6850 & 3.0722 & \textbf{1.5675} & 2.2179 & 2.1465 & 2.0769 \\
 & GEI(1)$\downarrow$ & 3.3227 & \textbf{1.8744} & 3.1159 & 2.8556 & 3.5190 & 2.2813 & 2.2657 & 2.1826 & \underline{2.1558} \\
 & GEI(2)$\downarrow$ & 20.2985 & \textbf{6.3348} & 20.3888 & 15.0595 & 24.1330 & 13.0966 & 8.6116 & \underline{8.0090} & 8.5757 \\
\bottomrule
\end{tabular}
\end{table*}

\begin{table*}[!h]
\centering
\caption{Overall results with SASRec as the backbone. The best value in each row is in \textbf{bold} and the second-best is \underline{underlined}. The backbone serves as an accuracy reference. $\downarrow$ indicates lower is better and $\uparrow$ higher is better.}
\label{tab:overall_sasrec}
\small
\setlength\tabcolsep{1.0pt}
\renewcommand{\arraystretch}{0.7}
\begin{tabular}{cc c cccc cccc}
\toprule
& & \multicolumn{1}{c}{Backbone} & \multicolumn{4}{c}{In-Processing} & \multicolumn{4}{c}{Post-Processing} \\
\cmidrule(lr){3-3} \cmidrule(lr){4-7} \cmidrule(lr){8-11}
Dataset & Metric & SASRec & SDRO & IPS & FOCF & FairDual & FairRec & CPFair & PMMF & ElasticRank \\
\midrule
\multirow{9}{*}{Steam} & NDCG@20 & \textbf{0.1813} & 0.1637 & 0.1631 & 0.1643 & \underline{0.1654} & 0.1604 & 0.1608 & 0.1627 & 0.1611 \\
 & HR@20 & \textbf{0.3855} & 0.3668 & 0.3358 & 0.3525 & 0.3671 & \underline{0.3833} & 0.3231 & 0.3126 & 0.3291 \\
 & MRR@20 & \textbf{0.1369} & 0.1193 & 0.1269 & 0.1233 & 0.1208 & 0.1079 & 0.1264 & \underline{0.1312} & 0.1258 \\
\cmidrule(lr){2-11}
 & Gini$\downarrow$ & 0.8348 & 0.7293 & 0.7352 & 0.7422 & 0.8007 & 0.7915 & \underline{0.6096} & \textbf{0.5484} & 0.6730 \\
 & Ent.$\uparrow$ & 3.1814 & 3.8952 & 3.8657 & 3.8435 & 3.4599 & 3.5311 & \underline{4.3850} & \textbf{4.5460} & 4.1619 \\
\cmidrule(lr){2-11}
 & GEI(0)$\downarrow$ & 15.9833 & 8.3375 & 10.0972 & \underline{2.5782} & 4.6266 & 5.9809 & 4.9677 & \textbf{2.2616} & 7.9748 \\
 & GEI(0.5)$\downarrow$ & 1.8939 & 1.3636 & 1.3703 & 1.2265 & 1.5684 & 1.4222 & \underline{0.8391} & \textbf{0.6162} & 1.0990 \\
 & GEI(1)$\downarrow$ & 1.5325 & 1.0377 & 1.0581 & 1.0736 & 1.3394 & 1.2901 & \underline{0.6982} & \textbf{0.5866} & 0.8529 \\
 & GEI(2)$\downarrow$ & 2.5138 & 1.2560 & 1.3998 & 1.5528 & 2.0828 & 2.0657 & \underline{0.9652} & \textbf{0.8911} & 1.0981 \\
\midrule
\multirow{9}{*}{Video \& Games} & NDCG@20 & \textbf{0.0601} & 0.0546 & 0.0544 & 0.0549 & 0.0544 & 0.0539 & 0.0539 & \underline{0.0550} & 0.0546 \\
 & HR@20 & \textbf{0.1390} & 0.1283 & 0.1281 & 0.1299 & 0.1292 & \underline{0.1375} & 0.1259 & 0.1185 & 0.1273 \\
 & MRR@20 & \textbf{0.0408} & 0.0367 & 0.0363 & 0.0369 & 0.0366 & 0.0329 & 0.0363 & \underline{0.0393} & 0.0368 \\
\cmidrule(lr){2-11}
 & Gini$\downarrow$ & 0.4496 & 0.5233 & 0.3937 & 0.7003 & 0.6757 & 0.3861 & 0.3863 & \underline{0.3359} & \textbf{0.3282} \\
 & Ent.$\uparrow$ & 6.8087 & 6.6334 & 6.9375 & 5.9959 & 6.0798 & 6.9449 & 6.9485 & \underline{7.0394} & \textbf{7.0520} \\
\cmidrule(lr){2-11}
 & GEI(0)$\downarrow$ & 0.3493 & 0.5957 & 0.2587 & 3.8738 & 2.8053 & 0.2434 & 0.2473 & \underline{0.1841} & \textbf{0.1776} \\
 & GEI(0.5)$\downarrow$ & 0.3366 & 0.4839 & 0.2526 & 1.0525 & 0.9347 & 0.2429 & 0.2433 & \underline{0.1830} & \textbf{0.1755} \\
 & GEI(1)$\downarrow$ & 0.3495 & 0.4710 & 0.2602 & 0.9129 & 0.8547 & 0.2551 & 0.2526 & \underline{0.1896} & \textbf{0.1809} \\
 & GEI(2)$\downarrow$ & 0.4684 & 0.5921 & 0.3246 & 1.2531 & 1.3063 & 0.3285 & 0.3242 & \underline{0.2318} & \textbf{0.2200} \\
\midrule
\multirow{9}{*}{BeerAdvocate} & NDCG@20 & \textbf{0.0472} & 0.0422 & 0.0426 & \underline{0.0432} & 0.0423 & 0.0418 & 0.0423 & 0.0428 & 0.0427 \\
 & HR@20 & \textbf{0.2345} & 0.2122 & 0.2187 & 0.2297 & 0.2295 & \underline{0.2301} & 0.2155 & 0.2072 & 0.2165 \\
 & MRR@20 & 0.0624 & 0.0551 & 0.0567 & \underline{0.0634} & \textbf{0.0648} & 0.0476 & 0.0556 & 0.0596 & 0.0560 \\
\cmidrule(lr){2-11}
 & Gini$\downarrow$ & 0.7614 & 0.7611 & 0.7183 & 0.9036 & 0.9416 & \textbf{0.6662} & 0.7020 & 0.6987 & \underline{0.6834} \\
 & Ent.$\uparrow$ & 7.5433 & 7.6505 & 7.8003 & 6.2241 & 5.6115 & \underline{7.9697} & 7.9442 & 7.9621 & \textbf{8.0252} \\
\cmidrule(lr){2-11}
 & GEI(0)$\downarrow$ & 2.0936 & 2.2365 & 1.4424 & 14.7312 & 19.1838 & \textbf{0.8928} & 1.5829 & 1.4764 & \underline{1.2221} \\
 & GEI(0.5)$\downarrow$ & 1.1627 & 1.1846 & 0.9969 & 2.2437 & 2.6162 & \textbf{0.8221} & 0.9555 & 0.9453 & \underline{0.8853} \\
 & GEI(1)$\downarrow$ & 1.2282 & 1.1538 & 1.0500 & 2.1426 & 2.5672 & \underline{0.9326} & 0.9503 & 0.9379 & \textbf{0.8941} \\
 & GEI(2)$\downarrow$ & 2.6469 & 1.9511 & 2.1624 & 6.1172 & 9.0033 & 1.9523 & 1.6628 & \underline{1.6218} & \textbf{1.5610} \\
\bottomrule
\end{tabular}
\end{table*}

\subsection{Empirical Consistency and Complementary Diagnostic Value of GEI (RQ1)}
\label{sec:rq1}
We first examine the empirical relationship between GEI and conventional provider exposure fairness metrics. Tables~\ref{tab:overall_lightgcn} and~\ref{tab:overall_sasrec} report the results across three datasets and two recommendation backbones. Overall, GEI and conventional metrics exhibit substantial consistency in their fairness evaluations. In particular, on Steam and Video \& Games, the methods achieving the lowest Gini coefficients also attain the lowest GEI values across all considered sensitivity settings for both backbones. Moreover, fairness-aware methods that improve conventional fairness metrics generally exhibit reduced GEI values. These observations suggest that GEI broadly aligns with conventional metrics in assessing provider exposure fairness. Notably, GEI(1) is mathematically related to Shannon entropy, so their agreement is expected rather than an independent empirical finding.

However, this consistency does not hold uniformly across methods and sensitivity settings. For example, on Video \& Games with LightGCN, FOCF reduces the Gini coefficient from 0.9178 to 0.8943 and increases Shannon entropy from 3.8069 to 4.3682, indicating improved fairness under conventional metrics. Nevertheless, GEI(0) increases from 9.4704 to 13.2171, while GEI(0.5), GEI(1), and GEI(2) all decrease. Since GEI(0) places greater emphasis on the lower end of the exposure distribution, this divergence suggests that improvements under conventional metrics can coexist with increased inequality from a lower-tail-sensitive perspective. Thus, conventional fairness improvements do not necessarily imply uniform improvements across different regions of the provider exposure distribution.

The method-level comparisons further illustrate the complementary value of GEI. On BeerAdvocate, FairRec achieves the lowest Gini coefficient under both backbones, whereas GEI(2) identifies SDRO as the best-performing method with LightGCN and ElasticRank with SASRec. These differences indicate that the preferred fairness intervention can depend on which part of the exposure distribution receives greater emphasis. Consequently, methods that appear favorable under conventional metrics may not perform equally well under every GEI sensitivity setting.

Overall, the results show that GEI is broadly consistent with conventional provider exposure fairness metrics while providing additional diagnostic information through its sensitivity-dependent evaluation. Its value lies not in replacing existing metrics, but in revealing distributional disparities that may be obscured by aggregate fairness scores. These findings motivate a more systematic investigation of sensitivity-dependent method rankings in RQ2 and the decomposition of provider exposure inequality in RQ3.

\subsection{Sensitivity-Dependent Comparisons and Ranking Reversals (RQ2)}
\label{sec:rq2}
\begin{figure*}[!h]
\centering
\includegraphics[width=\textwidth]{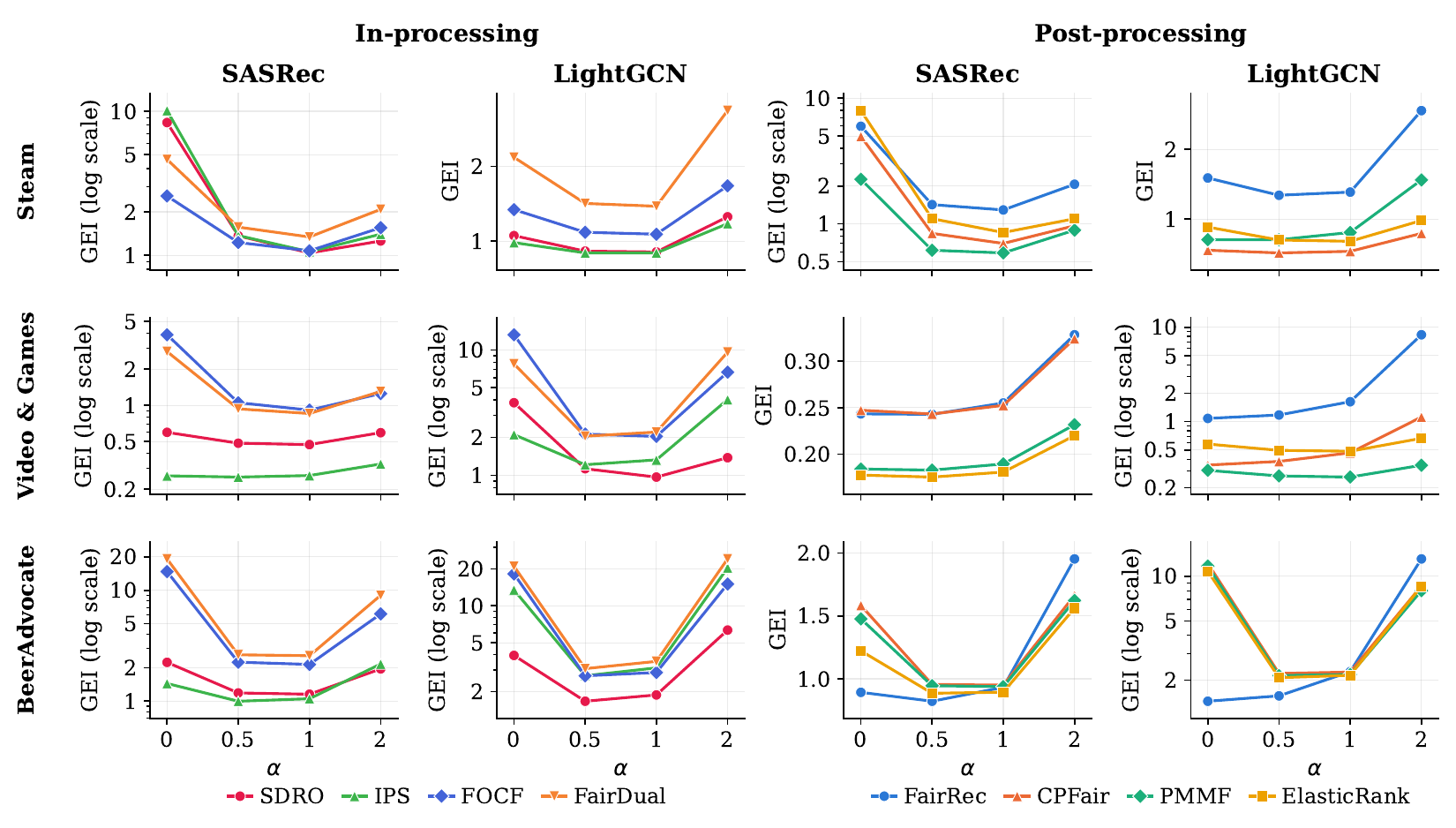}
\caption{Inequality profiles at $\alpha\in\{0,0.5,1,2\}$ on Steam, Video \& Games, and BeerAdvocate (rows). The left and right column pairs show in- and post-processing methods, respectively. Each pair shows SASRec followed by LightGCN. Logarithmic axes are marked. Compare methods at fixed $\alpha$ within each panel.}
\label{fig:profiles_combined}
\end{figure*}

We next investigate whether the relative fairness rankings of different methods remain stable as the sensitivity parameter $\alpha$ changes. For each method, we organize its GEI values over $\mathcal{S}=\{0,0.5,1,2\}$ into an \emph{inequality profile}. Figure~\ref{fig:profiles_combined} presents the profiles of the in- and post-processing methods. At a fixed sensitivity setting, lower GEI indicates better provider exposure fairness.

Two main observations emerge. First, some methods maintain their leading positions across sensitivity settings. Among the post-processing methods, PMMF consistently achieves the lowest GEI on Steam with SASRec and on Video \& Games with LightGCN. Similarly, CPFair remains the best-performing method on Steam with LightGCN, while ElasticRank maintains its advantage on Video \& Games with SASRec. Stable leading positions also appear among the in-processing methods. For example, IPS consistently performs best on Video \& Games with SASRec and Steam with LightGCN, whereas SDRO achieves the lowest GEI on BeerAdvocate with LightGCN. These results indicate that the identity of the best-performing method can remain stable across different sensitivity settings, even when the complete rankings of other methods may change.

Second, several methods exhibit ranking reversals as $\alpha$ changes. Among the in-processing methods, FOCF achieves the lowest GEI at $\alpha=0$ and $0.5$ on Steam with SASRec, whereas SDRO performs best at $\alpha=1$ and $2$. On Video \& Games with LightGCN, IPS leads at $\alpha=0$, but SDRO becomes the best-performing method at higher sensitivities. Ranking reversals also occur among the post-processing methods. On BeerAdvocate with SASRec, FairRec performs best at $\alpha=0$ and $0.5$, whereas ElasticRank takes the lead at $\alpha=1$ and $2$. An even more pronounced transition appears on BeerAdvocate with LightGCN, where FairRec achieves the lowest GEI at $\alpha=0$ and $0.5$, ElasticRank at $\alpha=1$, and PMMF at $\alpha=2$. These reversals demonstrate that the preferred fairness method can change depending on which region of the provider exposure distribution receives greater emphasis.

The inequality profiles further reveal sensitivity-dependent differences in the relative performance of individual methods. For example, on Steam with SASRec, IPS exhibits substantially higher GEI(0) than FOCF, whereas their GEI(1) values are nearly identical, with IPS performing slightly better. This contrast indicates that methods with comparable fairness under one sensitivity setting can exhibit considerably different behavior when greater emphasis is placed on less-exposed providers. Consequently, evaluating fairness at a single fixed value of $\alpha$ may overlook important differences in how methods distribute provider exposure.

Overall, the results show that the relative effectiveness of fairness-aware methods can be either stable or sensitivity-dependent. Multi-$\alpha$ analysis identifies methods whose leading positions persist across evaluation perspectives while revealing ranking reversals and distributional differences that a single fixed-sensitivity score would overlook. These findings motivate the subsequent decomposition analysis to investigate the structural sources of provider exposure inequality.

\begin{figure*}[!h]
\centering
\includegraphics[width=\textwidth]{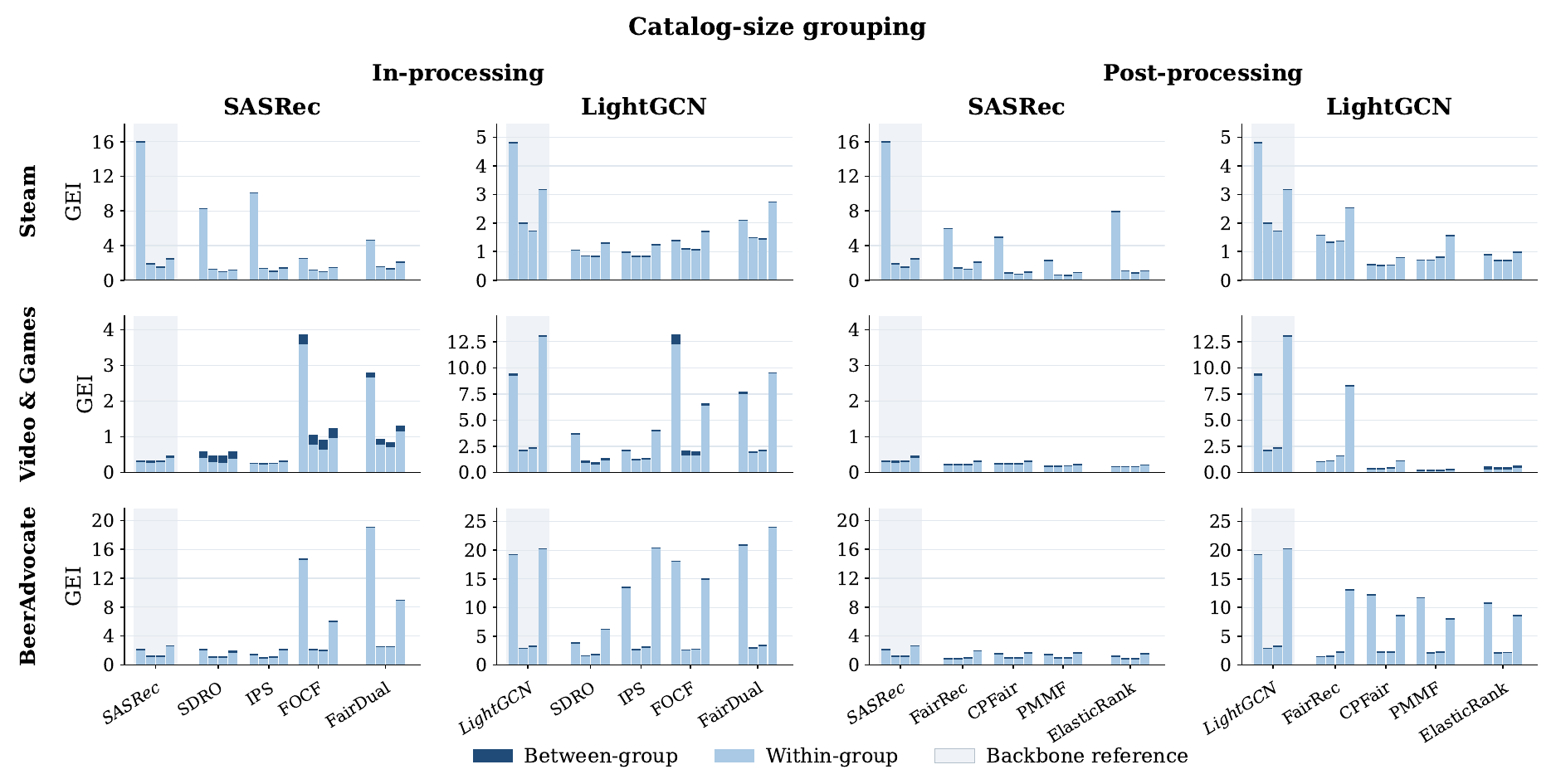}
\caption{GEI decomposition under catalog-size grouping. Rows show datasets, and the left and right column pairs show in- and post-processing methods, respectively. Each pair shows SASRec followed by LightGCN. The bars within each method are ordered by $\alpha=0,0.5,1,2$ and split into within- and between-group components.}
\label{fig:decomposition_size_combined}
\end{figure*}

\subsection{Structural Attribution (RQ3)}
\label{sec:rq3}
The analysis in RQ2 shows that fairness comparisons can change with the sensitivity setting, but it does not reveal how exposure unfairness is distributed within and between provider groups. We therefore use the additive decomposition of GEI to separate total unfairness into between-group and within-group components. We consider two provider-grouping attributes, catalog size and interaction frequency, and examine how structural attribution varies with both the grouping attribute and the sensitivity parameter $\alpha$.
\subsubsection{Structural Attribution across Provider Groupings}
\begin{figure*}[!t]
\centering
\includegraphics[width=\textwidth]{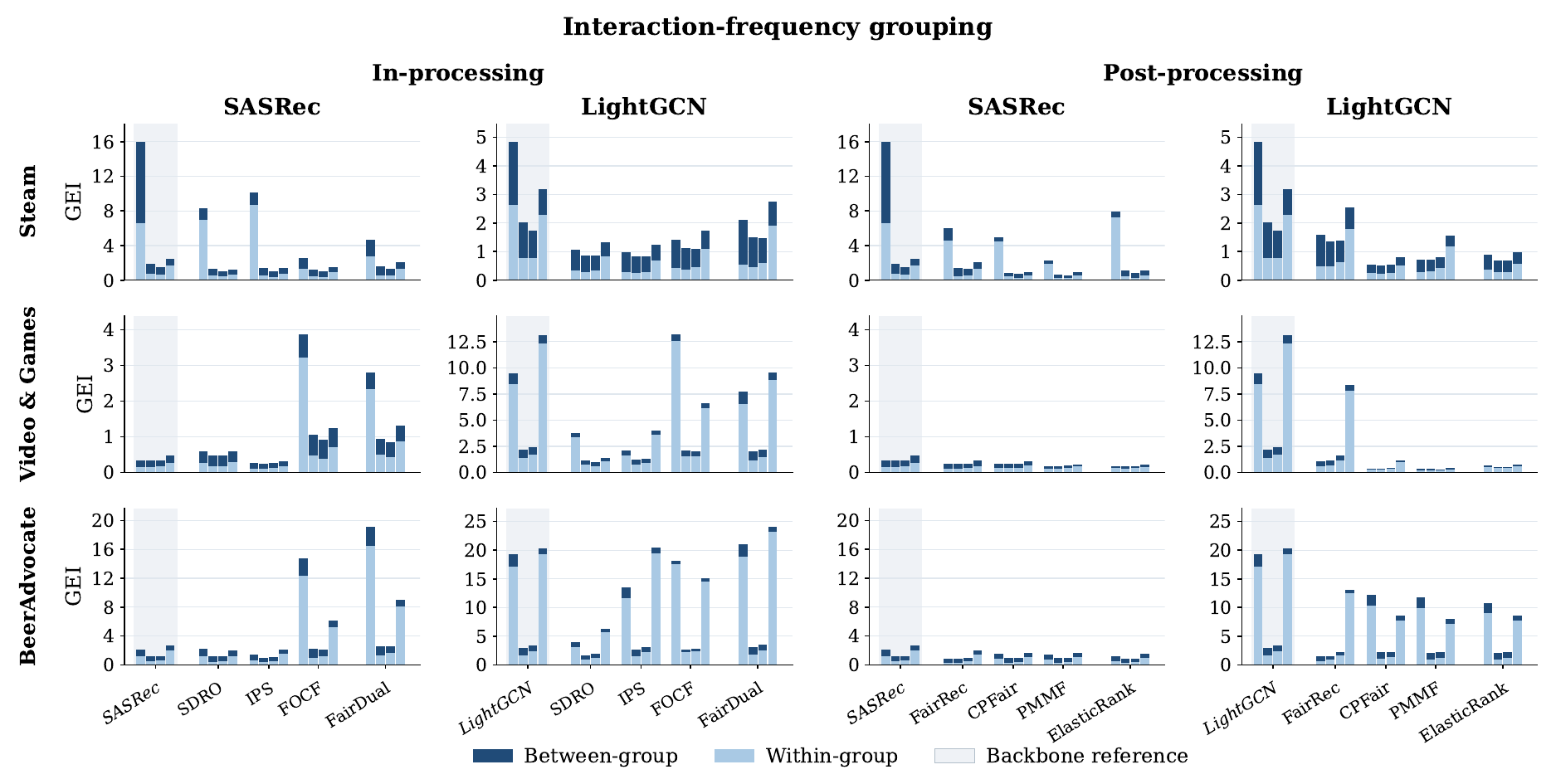}
\caption{GEI decomposition under interaction-frequency grouping, using the same panel and bar order as Figure~\ref{fig:decomposition_size_combined}. Each bar separates within- and between-group GEI.}
\label{fig:decomposition_freq_combined}
\end{figure*}

Figure~\ref{fig:decomposition_size_combined} presents the GEI decomposition under catalog-size grouping, while Figure~\ref{fig:decomposition_freq_combined} shows the results under interaction-frequency grouping. The two grouping attributes reveal different structural patterns. Under catalog-size grouping, within-group unfairness dominates in most settings, particularly on Steam and BeerAdvocate. In contrast, interaction-frequency grouping often attributes a substantially larger share of total unfairness to between-group differences, especially on Steam. This suggests that exposure inequality is not necessarily concentrated between providers of different catalog sizes, but may be more strongly associated with differences in their historical interaction frequencies.


The decomposition also reveals differences in how fairness-aware methods affect the two components. For example, on Steam with SASRec, FOCF substantially reduces within-group unfairness under catalog-size grouping, particularly at $\alpha=0$. Moreover, on Video \& Games with LightGCN, CPFair and PMMF achieve comparable total GEI at $\alpha=0$ but exhibit different structural compositions under catalog-size grouping. CPFair's remaining unfairness is primarily concentrated within groups, whereas PMMF shows more comparable between-group and within-group contributions. These findings indicate that improvements in total GEI do not necessarily imply simultaneous reductions in both components, and similar overall fairness scores can conceal different patterns of exposure inequality.

\begin{figure*}[!t]
\centering
\includegraphics[width=\textwidth]{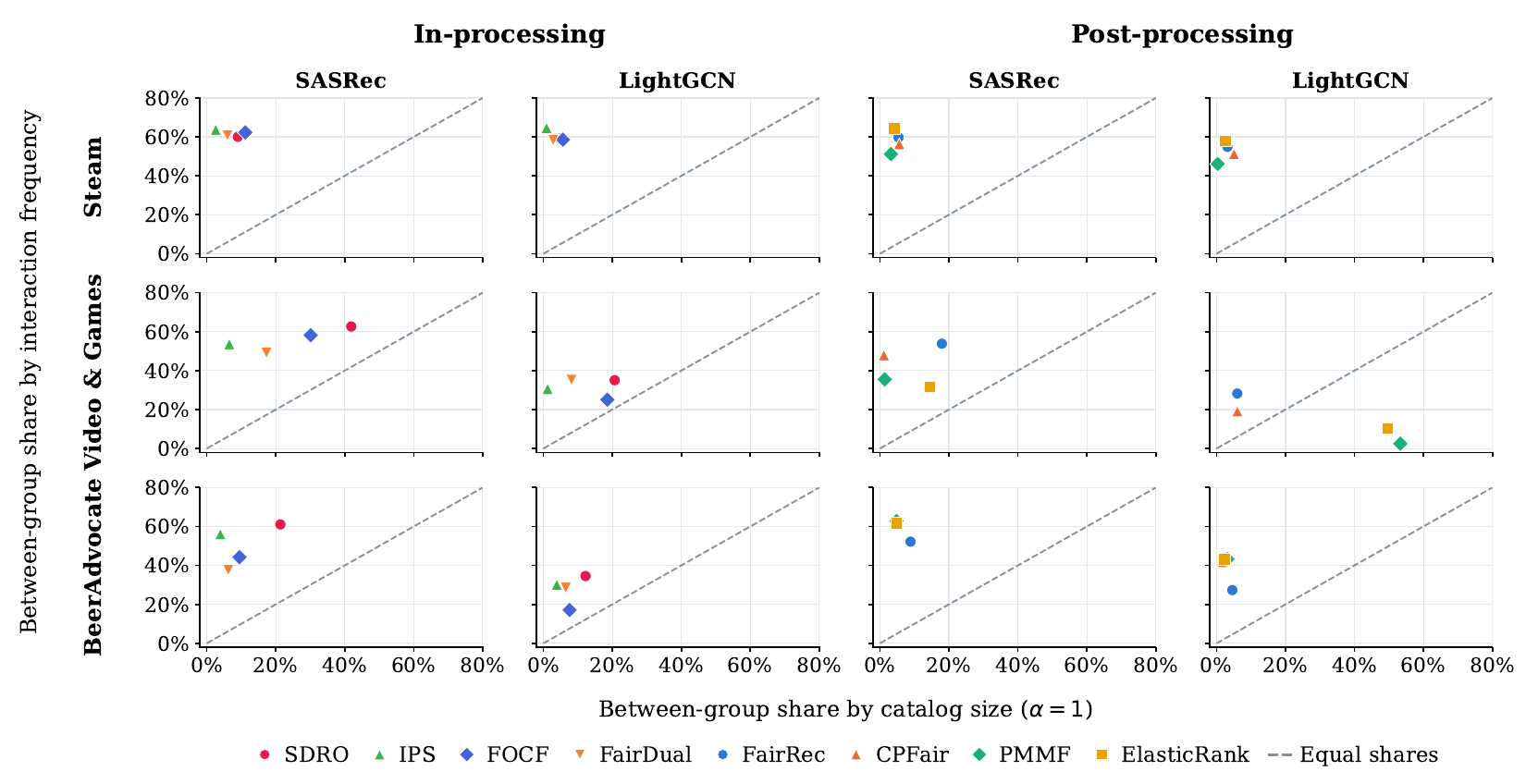}
\caption{Between-group GEI shares at $\alpha=1$ under catalog-size grouping (horizontal axis) and interaction-frequency grouping (vertical axis). Each point represents a method. The left and right column pairs show in- and post-processing methods, respectively. The dashed diagonal indicates equal shares.}
\label{fig:rho_size_vs_freq_combined}
\end{figure*}

Figure~\ref{fig:rho_size_vs_freq_combined} further compares the between-group shares under the two grouping attributes at $\alpha=1$ for in- and post-processing methods. On Steam, most methods exhibit substantially higher between-group shares under interaction-frequency grouping than under catalog-size grouping, regardless of the backbone. However, this pattern does not hold universally. For example, on Video \& Games with LightGCN, PMMF and ElasticRank exhibit larger between-group shares under catalog-size grouping. These contrasting results demonstrate that structural attribution depends on the grouping attribute, and neither grouping attribute consistently yields a larger between-group share across all settings.

Overall, these findings highlight the importance of considering different provider-grouping attributes when interpreting exposure unfairness. Although the total GEI remains unchanged for a given recommendation result, its decomposition can reveal substantially different contributions from between-group and within-group inequality. Thus, structural attribution provides complementary information about the distribution of unfairness that cannot be inferred from the overall GEI value alone.
\begin{figure*}[!h]
\centering
\includegraphics[width=0.95\textwidth]{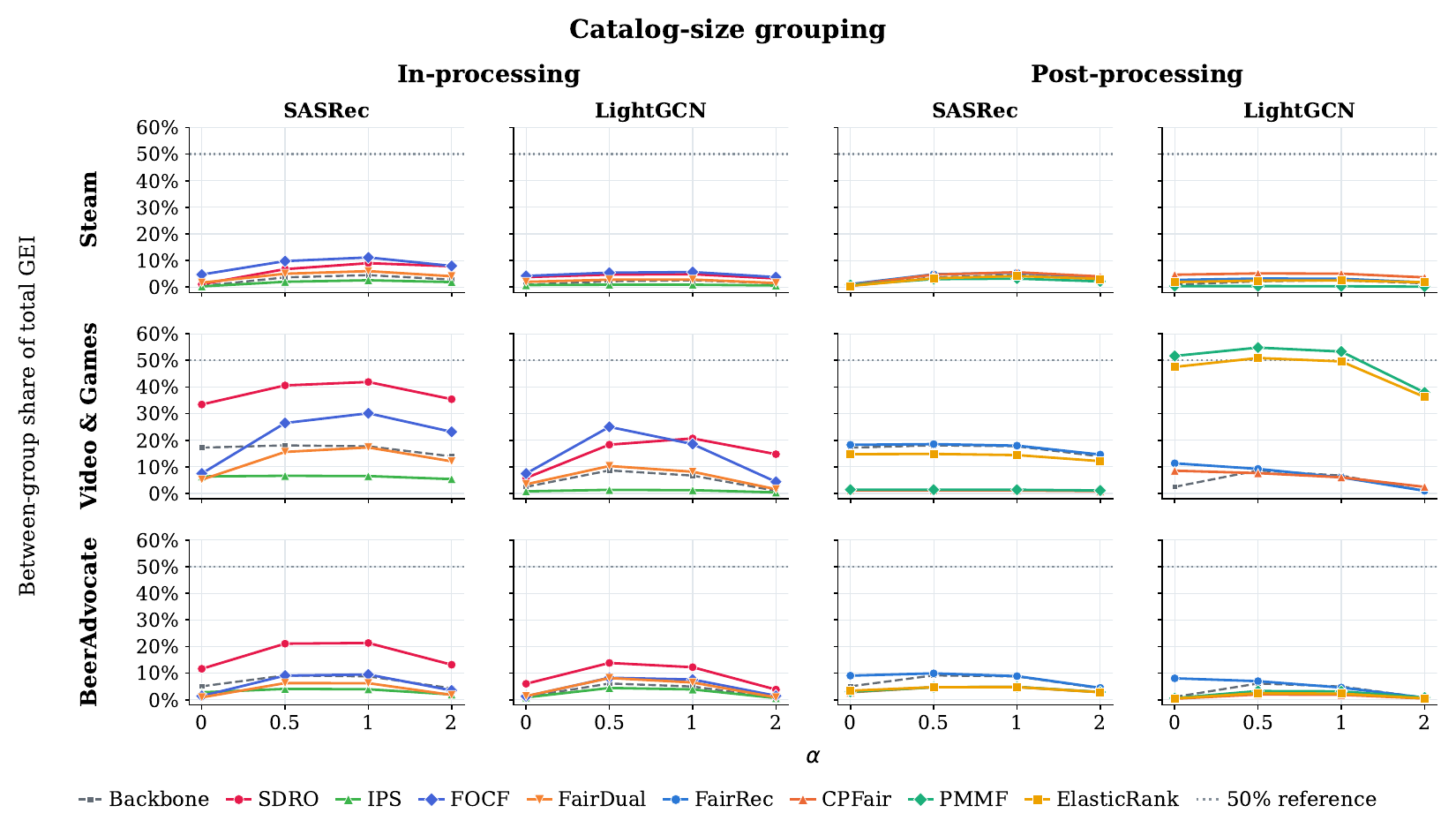}
\caption{Between-group share $\rho_n(\alpha)$ of total GEI under catalog-size grouping. The left and right column pairs show in- and post-processing methods, respectively. The dotted line marks a 50\% share.}
\label{fig:rho_profile_size_post}
\end{figure*}

\begin{figure*}[!h]
\centering
\includegraphics[width=0.95\textwidth]{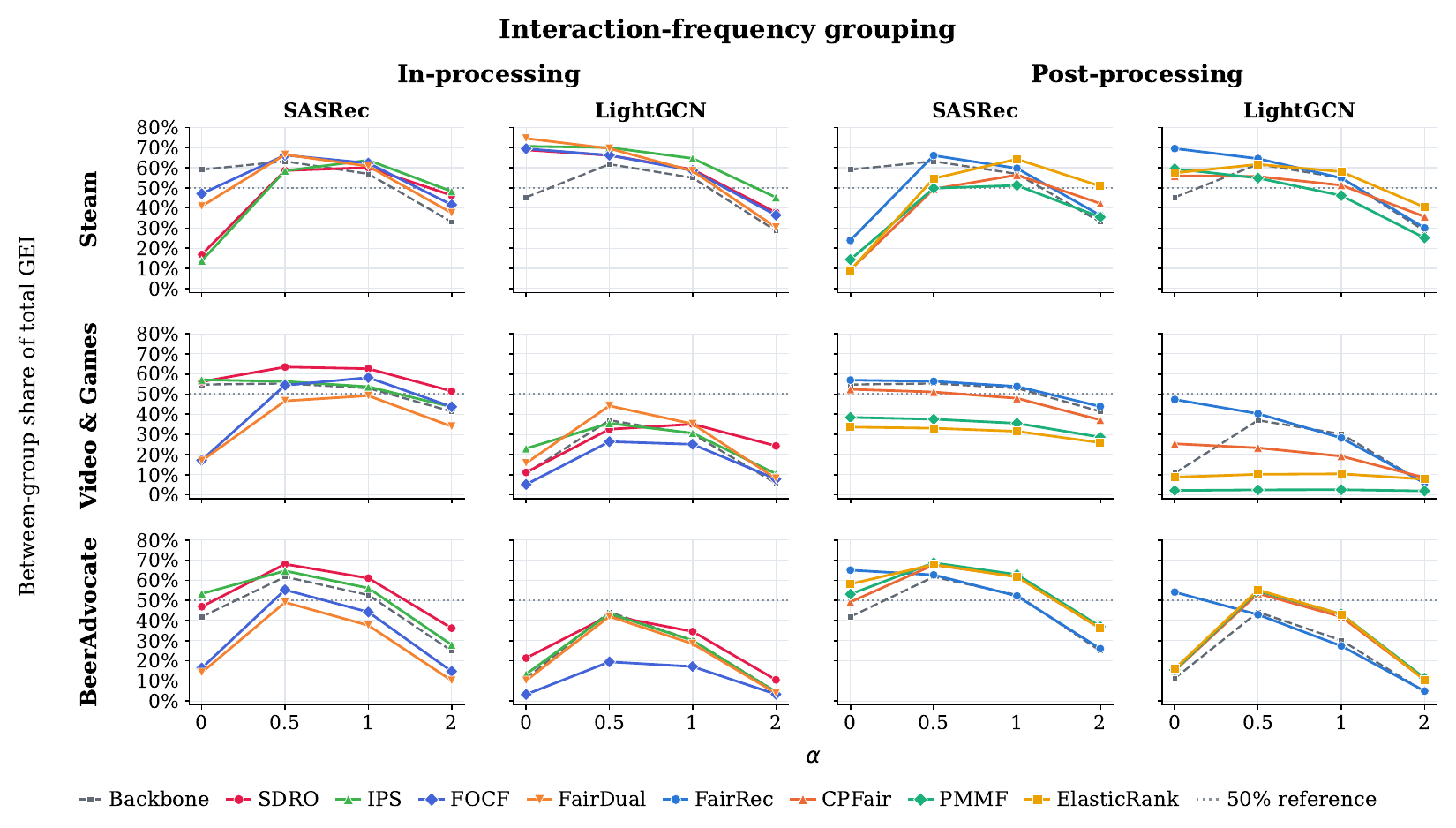}
\caption{Between-group share $\rho_n(\alpha)$ of total GEI under interaction-frequency grouping. The left and right column pairs show in- and post-processing methods, respectively. The dotted line marks a 50\% share.}
\label{fig:rho_profile_freq_combined}
\end{figure*}

\subsubsection{Sensitivity-Dependent Structural Attribution}
We further examine how the relative contributions of between-group and within-group unfairness change with the sensitivity parameter $\alpha$. Figure~\ref{fig:rho_profile_size_post} presents the diagnostic-ratio profiles under catalog-size grouping, while Figure~\ref{fig:rho_profile_freq_combined} shows the corresponding profiles under interaction-frequency grouping. These profiles reveal how the between-group share of total GEI varies across sensitivity settings.

Under catalog-size grouping, the between-group share remains relatively small across sensitivity settings in most cases, particularly on Steam and BeerAdvocate. However, some methods exhibit more pronounced sensitivity-dependent changes. For example, on Video \& Games with LightGCN, PMMF and ElasticRank maintain relatively high between-group shares at lower and intermediate sensitivities, but these shares decline at $\alpha=2$. This indicates that the relative contributions of between-group and within-group unfairness can change with evaluation sensitivity even when the grouping attribute remains fixed.

Under interaction-frequency grouping, the between-group share is generally more substantial, especially on Steam, and its variation with $\alpha$ becomes more pronounced. On Steam with SASRec, several methods exhibit increasing between-group shares at lower sensitivities, followed by a decline at higher sensitivities. A similar non-monotonic pattern appears on BeerAdvocate with SASRec, where many methods reach their highest between-group shares at intermediate sensitivity settings. These results show that the structural attribution of exposure unfairness does not necessarily change monotonically with $\alpha$.

The sensitivity patterns also differ across methods under the same grouping attribute. For example, on Video \& Games with LightGCN under interaction-frequency grouping, PMMF maintains a consistently small between-group share, whereas FairRec exhibits a substantially larger share that decreases as $\alpha$ increases. Such differences indicate that the relative importance of between-group and within-group unfairness depends on both the evaluated method and the sensitivity setting. Importantly, changes in the between-group share reflect changes in the relative contributions of the two components rather than necessarily indicating an improvement or deterioration in overall fairness.

\subsection{Design Implications (RQ4)}
\label{sec:rq4}

We illustrate how the diagnostic findings from RQ1--RQ3 can support fairness-aware method selection through an end-to-end case study. We then discuss the broader implications for diagnosis-informed method selection.

\subsubsection{Case Study on BeerAdvocate with SASRec}
\label{sec:rq4_case}
\begin{figure*}[!h]
\centering
\includegraphics[width=\textwidth]{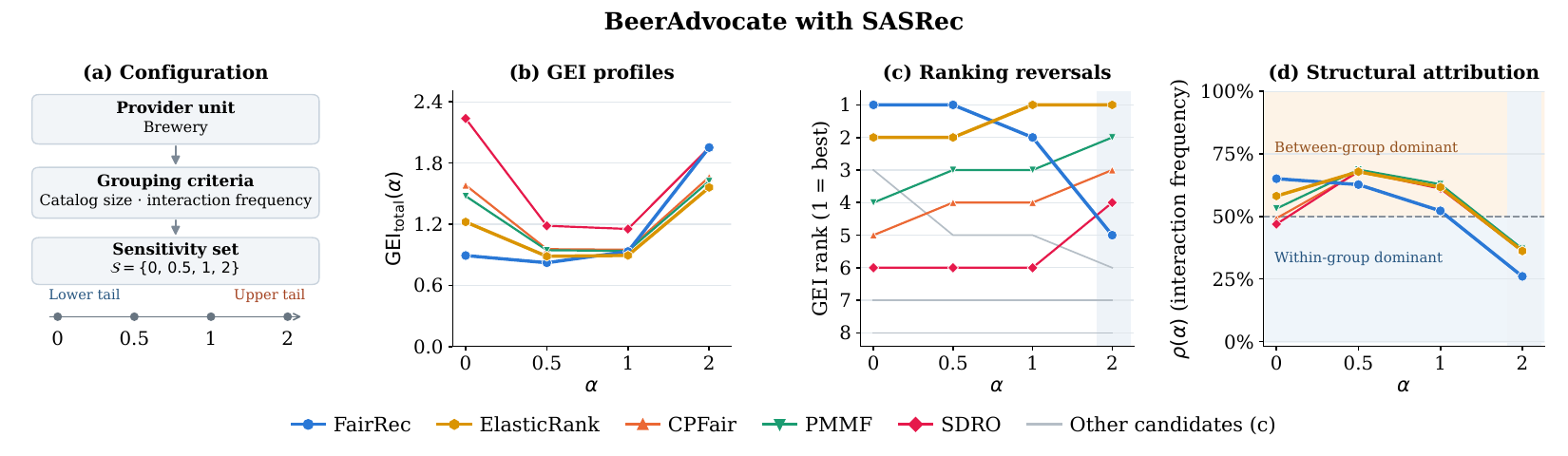}
\caption{Case study on BeerAdvocate with SASRec. The four panels illustrate (a) diagnostic configuration, (b) GEI profiles across sensitivity settings, (c) sensitivity-dependent method rankings, and (d) diagnostic-ratio profiles under interaction-frequency grouping.}
\label{fig:case_study}
\end{figure*}

We consider BeerAdvocate with SASRec, where the preferred fairness-aware method changes with the evaluation sensitivity. Figure~\ref{fig:case_study} illustrates the four-stage diagnostic process, highlighting five representative methods while retaining all eight fairness-aware methods for ranking comparisons. Results for the complete candidate set are reported in Table~\ref{tab:overall_sasrec}.

In Stage~1, breweries are treated as providers and grouped by catalog size and interaction frequency. We evaluate the sensitivity set $\mathcal{S}=\{0,0.5,1,2\}$. To illustrate a concern about excessive exposure concentration among highly exposed providers, we adopt $\alpha=2$ as the primary evaluation setting. We consider FairRec, ElasticRank, CPFair, PMMF, and SDRO representative methods.

In Stage~2, inequality profiling reveals different fairness patterns across sensitivity settings. FairRec achieves the lowest GEI at lower sensitivities, indicating an advantage when the evaluation emphasizes underexposed providers. In contrast, ElasticRank performs better at higher sensitivities, suggesting stronger performance in limiting exposure concentration among highly exposed providers. These differences indicate that the relative effectiveness of fairness-aware methods depends on the fairness concern emphasized by the evaluation.

In Stage~3, the method rankings further demonstrate this sensitivity dependence. FairRec ranks first at $\alpha=0$ and $\alpha=0.5$, whereas ElasticRank becomes the best-performing method at $\alpha=1$ and $\alpha=2$. Conventional fairness metrics also yield conflicting preferences: the Gini coefficient favors FairRec, whereas Shannon entropy favors ElasticRank. Given the stated concern about upper-tail exposure concentration, ElasticRank is the preferred candidate under $\alpha=2$. Under a different concern emphasizing severe underexposure, FairRec would be a more suitable choice.

In Stage~4, structural attribution provides further insights into the remaining unfairness of the selected method. As shown in Figure~\ref{fig:case_study}(d), the between-group share under interaction-frequency grouping varies with $\alpha$ and decreases at higher sensitivities. At $\alpha=2$, within-group unfairness accounts for the larger share of total GEI. Under catalog-size grouping, the between-group contribution is even smaller, as shown in Figure~\ref{fig:rho_profile_size_post}. These results suggest that the remaining unfairness is primarily concentrated within provider groups under the stated evaluation setting, although the relative contribution of between-group differences depends on the grouping attribute.

Taken together, the case study identifies ElasticRank as the preferred candidate for the stated upper-tail fairness concern. The selection differs from that suggested by the Gini coefficient and changes when the evaluation emphasizes underexposed providers. Structural attribution further reveals where the remaining unfairness is concentrated. This case demonstrates how sensitivity-dependent comparisons and structural decomposition can jointly support fairness-aware method selection and diagnosis.

\subsubsection{From Diagnostic Evidence to Intervention Implications}
\label{sec:rq4_rules}
The case study suggests three general implications for method selection. First, the evaluation sensitivity should reflect the fairness concern of the application, as different sensitivity settings can favor different methods. Second, method rankings should be examined across sensitivity settings rather than assumed to be fixed. When ranking reversals occur, the preferred method should be selected according to the sensitivity setting aligned with the stated fairness concern. Third, structural attribution provides additional information about the sources of remaining unfairness. Because this attribution depends on the provider-grouping attribute, multiple meaningful groupings should be considered when interpreting the results and identifying potential intervention priorities.

Overall, the results answer RQ4 by demonstrating how inequality profiling, sensitivity-dependent ranking analysis, and structural attribution can be integrated into a diagnosis-informed method selection process. Rather than prescribing a universally optimal fairness method, the proposed framework helps evaluators select methods according to their fairness concerns and better understand the structure of the remaining provider exposure inequality.

\subsection{Robustness to Accuracy Targets}
\begin{table*}[!h]

\centering
\caption{Overall results on Steam at the $80\%$ validation NDCG@20 retention target. The best value in each column is in \textbf{bold} and the second-best is \underline{underlined}. The backbone serves as an accuracy reference. $\downarrow$ indicates lower is better and $\uparrow$ higher is better.}
\label{tab:steam_ndcg80}
\small
\setlength\tabcolsep{1.5pt}
\renewcommand{\arraystretch}{0.7}

\begin{tabular}{l ccc cccccc}
\toprule
\multicolumn{10}{c}{\textit{LightGCN}} \\
\midrule
Method & NDCG@20 & HR@20 & MRR@20
& Gini$\downarrow$ & Ent.$\uparrow$
& GEI(0)$\downarrow$
& GEI(0.5)$\downarrow$
& GEI(1)$\downarrow$
& GEI(2)$\downarrow$ \\
\midrule

Baseline
& \textbf{0.1591} & \textbf{0.3614} & \textbf{0.1126}
& 0.8677 & 2.8749
& 4.8393 & 2.0100 & 1.7449 & 3.1883 \\
\midrule

SDRO
& \underline{0.1302} & 0.2949 & 0.0931
& 0.6496 & 4.2171
& 1.0377 & 0.8306 & 0.8146 & 1.2525 \\

IPS
& 0.1270 & 0.2581 & 0.0990
& 0.6784 & 4.1681
& 1.0370 & 0.8698 & 0.8486 & 1.1795 \\

FOCF
& 0.1277 & 0.3027 & 0.0881
& 0.6698 & 4.1891
& 0.9949 & 0.8406 & 0.8340 & 1.2442 \\

FairDual
& 0.1253 & 0.2801 & 0.0917
& 0.7334 & 3.8010
& 1.1741 & 1.0370 & 1.1030 & 2.0239 \\
\midrule

FairRec
& 0.1253 & \underline{0.3560} & 0.0681
& 0.7267 & 3.7723
& 1.1156 & 1.0379 & 1.1229 & 2.0191 \\

CPFair
& 0.1256 & 0.2340 & \underline{0.1054}
& \textbf{0.3872} & \textbf{4.9837}
& \textbf{0.2569} & \textbf{0.2606}
& \textbf{0.2832} & \textbf{0.4084} \\

PMMF
& 0.1264 & 0.2493 & 0.1009
& \underline{0.4700} & 4.7757
& \underline{0.3743} & \underline{0.3839}
& 0.4274 & 0.6787 \\

ElasticRank
& 0.1256 & 0.2337 & \underline{0.1054}
& 0.4934 & \underline{4.7830}
& 0.5687 & 0.4542
& \underline{0.4223} & \underline{0.5101} \\

\midrule
\multicolumn{10}{c}{\textit{SASRec}} \\
\midrule

Method & NDCG@20 & HR@20 & MRR@20
& Gini$\downarrow$ & Ent.$\uparrow$
& GEI(0)$\downarrow$
& GEI(0.5)$\downarrow$
& GEI(1)$\downarrow$
& GEI(2)$\downarrow$ \\
\midrule

Baseline
& \textbf{0.1813} & \textbf{0.3855} & \textbf{0.1369}
& 0.8348 & 3.1814
& 15.9833 & 1.8939 & 1.5325 & 2.5138 \\
\midrule

SDRO
& 0.1483 & 0.3424 & 0.1064
& 0.6574 & 4.2123
& 7.0585 & 1.1017 & 0.8179 & 0.8693 \\

IPS
& 0.1483 & 0.3025 & 0.1163
& 0.6634 & 4.1928
& 7.0458 & 1.0849 & 0.8314 & 1.0144 \\

FOCF
& \underline{0.1489} & 0.3175 & 0.1128
& 0.6586 & 4.2468
& \underline{1.6962} & 0.9090 & 0.7940 & 0.9951 \\

FairDual
& 0.1454 & 0.3454 & 0.1005
& 0.7788 & 3.5770
& 6.2951 & 1.4472 & 1.2582 & 1.9604 \\
\midrule

FairRec
& 0.1421 & \underline{0.3798} & 0.0839
& 0.7512 & 3.7901
& 5.6654 & 1.2256 & 1.1106 & 1.6966 \\

CPFair
& 0.1433 & 0.2901 & 0.1126
& \underline{0.4914} & \underline{4.7207}
& 1.8082 & \underline{0.5215}
& \underline{0.4655} & \underline{0.6642} \\

PMMF
& 0.1478 & 0.2707 & \underline{0.1233}
& \textbf{0.4888} & \textbf{4.7404}
& \textbf{1.1603} & \textbf{0.4831}
& \textbf{0.4519} & \textbf{0.6339} \\

ElasticRank
& 0.1427 & 0.2978 & 0.1106
& 0.5846 & 4.4594
& 5.5043 & 0.8463 & 0.6467 & 0.7912 \\

\bottomrule
\end{tabular}
\end{table*}

We repeat the comparison on Steam with a lower NDCG retention target to examine how relaxing the accuracy requirement affects the fairness results. Table~\ref{tab:steam_ndcg80} shows that test NDCG decreases, while GEI generally decreases when the same method is compared at the same sensitivity setting. The leading methods remain the same as in the main experiment. CPFair with LightGCN and PMMF with SASRec achieve the lowest GEI across all evaluated sensitivity settings.

\begin{figure*}[!h]
\centering
\includegraphics[width=\textwidth]{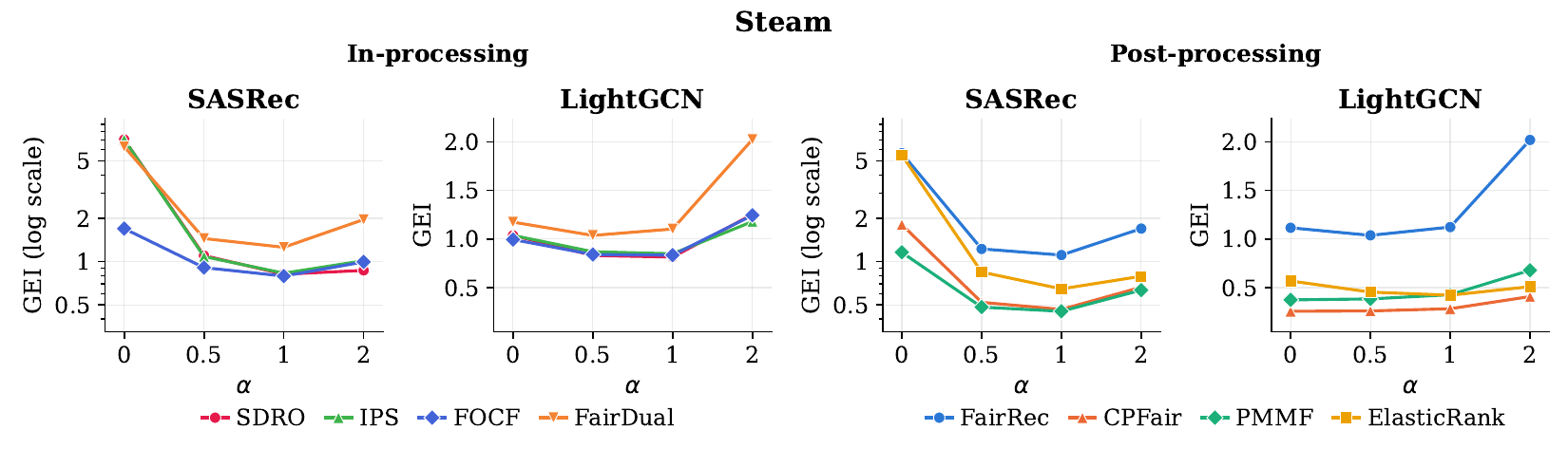}
\caption{Inequality profiles on Steam under the lower NDCG retention target. The left two panels show in-processing methods, and the right two show post-processing methods. Within each pair, SASRec is shown first on a logarithmic scale, followed by LightGCN on a linear scale. Each curve connects the GEI values of one method across sensitivity settings. Methods are compared at the same value of $\alpha$ within each panel.}
\label{fig:steam_gei_profiles_ndcg80}
\end{figure*}

Figure~\ref{fig:steam_gei_profiles_ndcg80} shows that ranking reversals still occur under the lower target. Among the in-processing methods with SASRec, FOCF leads when the evaluation emphasizes underexposed providers, and SDRO takes the lead when the focus shifts to exposure concentration among highly exposed providers. Among the post-processing methods with LightGCN, PMMF has lower GEI than ElasticRank at lower sensitivities, and ElasticRank moves ahead at higher sensitivities. These reversals show that the choice between methods still depends on which part of the exposure distribution is emphasized.

\subsection{Stability of Ranking Reversals}

We investigate whether the ranking reversals observed in RQ2 remain consistent across independent runs and extend the analysis to comparisons between in-processing and post-processing methods. This analysis is conducted on Steam with SASRec as the backbone. We focus on $\alpha=0$ and $\alpha=2$ because their comparison captures most of the ranking reversals identified in RQ2 and represents the two ends of the evaluated sensitivity range, emphasizing underexposed and highly exposed providers, respectively. For any two methods $A$ and $B$, we define their GEI difference at sensitivity $\alpha$ as $\Delta_\alpha = \mathrm{GEI}_A(\alpha) - \mathrm{GEI}_B(\alpha)$. To avoid treating negligible numerical differences as ranking reversals, we define a numerical tie tolerance as $\tau_\alpha = \max(10^{-6}, 10^{-4}\max(|\mathrm{GEI}_A(\alpha)|, |\mathrm{GEI}_B(\alpha)|))$. A ranking reversal is identified only when the GEI differences at the two sensitivity settings have opposite signs and both exceed their corresponding tolerances.
\begin{table}[!h]
\centering
\caption{Cross-run stability of ranking reversals on Steam with SASRec. For $\Delta_0$ and $\Delta_2$, we report the mean $\pm$ sample standard deviation of the paired GEI differences across five runs. Agreement indicates the percentage of paired runs reproducing the mean ordering at both sensitivity settings.}
\label{tab:reversal_stability}
\scriptsize
\setlength{\tabcolsep}{2pt}
\renewcommand{\arraystretch}{1.}
\begin{tabular}{llrrr}
\toprule
Category & Method Pair & $\Delta_0$ (Mean $\pm$ Std) & $\Delta_2$ (Mean $\pm$ Std) & Agree. (\%) \\
\midrule
\multirow{5}{*}{Within-category}
& SDRO / FOCF
& $+5.7593 \pm 2.1611$
& $-0.2968 \pm 0.1576$ & 100 \\
& SDRO / FairDual
& $+3.7109 \pm 3.5155$
& $-0.8268 \pm 0.0708$ & 80 \\
& IPS / FOCF
& $+7.5190 \pm 1.1065$
& $-0.1530 \pm 0.1167$ & 80 \\
& IPS / FairDual
& $+5.4706 \pm 1.6459$
& $-0.6830 \pm 0.0683$ & 100 \\
& FairRec / ElasticRank
& $-1.9939 \pm 0.6220$
& $+0.9676 \pm 0.0801$ & 100 \\

\midrule
\multirow{7}{*}{Cross-category}
& SDRO / FairRec
& $+2.3566 \pm 2.0076$
& $-0.8097 \pm 0.0492$ & 80 \\
& IPS / FairRec
& $+4.1163 \pm 1.3577$
& $-0.6659 \pm 0.0945$ & 100 \\
& FOCF / CPFair
& $-2.3895 \pm 0.8964$
& $+0.5876 \pm 0.2015$ & 100 \\
& FOCF / ElasticRank
& $-5.3966 \pm 0.8325$
& $+0.4547 \pm 0.1609$ & 100 \\
& FairDual / FairRec
& $-1.3543 \pm 0.9259$
& $+0.0171 \pm 0.0798$ & 20 \\
& FairDual / CPFair
& $-0.3411 \pm 1.1078$
& $+1.1176 \pm 0.0547$ & 80 \\
& FairDual / ElasticRank
& $-3.3482 \pm 1.1340$
& $+0.9847 \pm 0.0153$ & 100 \\
\bottomrule
\end{tabular}
\end{table}

To assess the reliability of these reversals, we calculate the mean and sample standard deviation of the GEI differences across runs paired by random seed. We also measure the proportion of paired runs in which the same ranking reversal is reproduced. Table~\ref{tab:reversal_stability} reports these statistics. We first examine comparisons among in-processing methods and among post-processing methods, consistent with RQ2. Most ranking reversals are reproduced across all runs, while the remaining comparisons also show high agreement. For example, SDRO and FOCF, as well as FairRec and ElasticRank, consistently reverse their relative fairness rankings as $\alpha$ changes. These results support the high cross-run consistency of the ranking reversals observed among methods using the same intervention strategy.

We then consider comparisons between in-processing and post-processing methods. Most of these comparisons exhibit high agreement across independent runs. For example, IPS and FairRec, as well as FOCF and CPFair, consistently reverse their rankings across sensitivity settings. However, the reversal between FairDual and FairRec shows lower agreement across runs, with a small mean GEI difference relative to its cross-run variability. This contrast highlights how ranking-reversal stability varies across method pairs, particularly when their fairness scores are close.

Overall, most observed ranking reversals are consistently reproduced across independent runs, while a small number exhibit greater variability. By jointly examining GEI differences, their cross-run variability, and reversal agreement, we provide a more comprehensive assessment of ranking-reversal stability. These results demonstrate the cross-run reproducibility of most observed ranking reversals and the value of examining their stability beyond mean GEI scores.

\subsection{Zero Exposure and Smoothing Sensitivity}
We investigate how zero-exposure providers and the choice of smoothing parameter affect GEI(0) on Steam with SASRec, using a fixed random seed of 2025. Table~\ref{tab:smoothing_sensitivity} reports the number of zero-exposure providers and the corresponding GEI(0) values under different smoothing parameters.

\begin{table}[!h]
\centering
\caption{Zero-exposure provider counts and sensitivity of GEI(0) to the smoothing parameter $\eta$ on Steam with SASRec (seed 2025). Zero-exp. denotes the number of providers with zero exposure before smoothing.}
\label{tab:smoothing_sensitivity}
\scriptsize
\setlength{\tabcolsep}{5pt}
\renewcommand{\arraystretch}{1.}
\begin{tabular}{lrrrr}
\toprule
Method & Zero-exp. & \multicolumn{3}{c}{GEI(0)} \\
\cmidrule(lr){3-5}
& & $\eta=10^{-12}$ & $\eta=10^{-9}$ & $\eta=10^{-6}$ \\
\midrule
SASRec & 23 & 15.4369 & 11.6541 & 7.8712 \\
SDRO & 8 & 6.7588 & 5.4430 & 4.1271 \\
IPS & 10 & 7.8074 & 6.1627 & 4.5178 \\
FOCF & 0 & 1.7658 & 1.7658 & 1.7658 \\
FairDual & 1 & 3.3740 & 3.2095 & 3.0449 \\
FairRec & 7 & 5.7569 & 4.6056 & 3.4543 \\
CPFair & 7 & 5.1937 & 4.0424 & 2.8911 \\
PMMF & 2 & 1.9556 & 1.6266 & 1.2976 \\
ElasticRank & 10 & 7.3101 & 5.6654 & 4.0205 \\
\bottomrule
\end{tabular}
\end{table}

The SASRec backbone leaves more than half of the providers without exposure, whereas all fairness-aware methods reduce this number. Among these methods, IPS and ElasticRank have the most zero-exposure providers, while FOCF provides exposure to every provider. To isolate the effect of smoothing, we fix the provider exposure distributions and recompute GEI(0) under different values of $\eta$. As $\eta$ increases, GEI(0) decreases more substantially for methods with more zero-exposure providers, such as IPS and ElasticRank. This pattern is consistent with the logarithmic dependence of zero-exposure contributions on the smoothing parameter. In contrast, FOCF's GEI(0) remains unchanged at the reported precision. Notably, FOCF has lower GEI(0) than PMMF at the smallest smoothing parameter, but their relative ranking reverses as $\eta$ increases. These findings demonstrate that smoothing affects not only GEI(0) values but also fairness comparisons.

These results provide practical guidance for applying our diagnostic framework. When evaluating fairness at $\alpha=0$, the number of zero-exposure providers and the smoothing parameter should be reported explicitly, and the stability of key method comparisons should be examined under different smoothing settings. Moreover, GEI(0) should be interpreted alongside other sensitivity settings to distinguish the effects of smoothing from differences in the underlying exposure distributions. This allows the framework to provide a more comprehensive assessment of provider exposure fairness.

\section{Practical Scope and Limitations}
\label{Limitations}

The proposed framework is a diagnostic tool rather than a fairness mitigation algorithm. It can support method selection and identify potential intervention priorities, but it does not directly improve exposure fairness. The choice of fairness intervention still depends on the platform's fairness objectives, accuracy requirements, and operational constraints.

Structural decomposition requires a well-defined partition of providers into mutually exclusive groups. While overall GEI and sensitivity-based comparisons do not require group memberships, the between-group and within-group decomposition relies on a known group partition. When group memberships are unknown, the current decomposition cannot be directly applied. Different assignments of the same providers may lead to different between-group and within-group contributions, even when the overall GEI remains unchanged. One possible extension is to estimate group memberships from available provider attributes and examine the robustness of structural attribution across different plausible group assignments. However, the resulting attribution would depend on how these memberships are estimated and how uncertainty is handled. Addressing these challenges remains a direction for future work.

The diagnostic results also depend on how providers and their exposure are defined. In our experiments, we adopt a simplified preprocessing strategy that merges providers with very few items into a single auxiliary provider to reduce the number of sparsely represented provider units. However, this aggregation changes the granularity of fairness evaluation. It may conceal exposure disparities among small providers and affect overall GEI values, between-group and within-group decomposition results, and even the relative fairness rankings of recommendation methods. Although we apply the same provider mapping and aggregation rules to all evaluated methods to ensure consistent evaluation conditions, this does not guarantee that the fairness conclusions are unaffected by aggregation. Therefore, our experimental findings should be interpreted in light of the adopted provider definition and aggregation strategy.

The robustness of ranking reversals also warrants consideration. Our experiments on Steam show that ranking reversals across GEI sensitivity settings persist under different random seeds and accuracy retention targets, indicating that this phenomenon is not limited to a single experimental run or a particular accuracy constraint. However, variations in random seeds may affect the relative fairness rankings of individual methods, and not all method pairs exhibit equally stable ranking reversals. Furthermore, when some providers receive no exposure, GEI(0) becomes sensitive to the choice of smoothing parameter. Different smoothing settings can change GEI(0) values and may even reverse the relative fairness rankings of recommendation methods. Therefore, ranking-reversal findings should be interpreted with attention to variability across random seeds and, when zero-exposure providers are present, the influence of smoothing on fairness comparisons.

Finally, the current framework evaluates static exposure estimated from top-$K$ recommendation lists. Such exposure does not necessarily reflect actual user attention or long-term provider outcomes. In real-world platforms, exposure accumulates over repeated recommendation cycles and may influence user feedback, provider participation, and item supply. Extending the framework to dynamic recommendation settings that capture interactions among users, providers, and recommender systems is therefore an important direction for future work.

\section{Conclusion}
\label{conclu}
This paper introduces a GEI-based diagnostic framework for provider exposure fairness. The tunable sensitivity parameter allows evaluators to examine whether fairness conclusions remain stable across distributional perspectives, while additive decomposition separates total unfairness into between-group and within-group components. Based on these properties, we develop a four-stage protocol consisting of configuration, inequality profiling, ranking-reversal analysis, and structural attribution. Experiments across multiple datasets, recommendation backbones, and fairness-aware methods produce two main findings. First, method rankings can change or reverse as the sensitivity parameter varies, showing that a single fixed-sensitivity score may conceal important evaluation instability. Second, the relative contributions of between-group and within-group unfairness vary across methods, datasets, and provider partitions. Methods with similar total unfairness can therefore exhibit different structural compositions.

\bibliographystyle{ACM-Reference-Format}
\bibliography{sample-base}

\end{document}